\documentclass{article} 

\usepackage{mathtools,amsthm,amssymb}
\newtheorem{definition}{Definition}
\newtheorem{theorem}{Theorem}
\newtheorem{lemma}{Lemma}
\newtheorem{proposition}{Proposition}
\usepackage{subcaption}
\usepackage{graphicx}

\def\rrangle{\rangle\!\rangle}
\newcommand{\vecket}[1]{| #1 \rrangle}

\usepackage{tikz}
\usetikzlibrary{positioning, shapes.geometric, arrows.meta, calc}
\usetikzlibrary{quantikz}
\usetikzlibrary{decorations.pathreplacing, decorations.pathmorphing}
\tikzset{
    triangle/.style = {
        regular polygon,
        regular polygon sides = 3,
        inner sep = 0pt,
        minimum size = 1cm
    },
    triangle right/.style = {
        triangle,
        shape border rotate = 30
    },
    triangle left/.style = {
        triangle,
        shape border rotate = 330
    }
}

\usepackage{geometry}
\usepackage{setspace}
\usepackage{authblk}

\begin{document}

\title{Quantum Parallel Algorithm for Multivariate Trace Estimation with Applications}

\author[1,2]{Honghong Lin}
\affil[1]{Institute of Mathematics, Academy of Mathematics and Systems Science, Chinese Academy of Sciences, Beijing 100190, China}
\affil[2]{School of Mathematical Sciences, University of Chinese Academy of Sciences, Beijing 100049, China}
\author[1,3]{Yun Shang\footnote{Shangyun@amss.ac.cn}}
\affil[3]{State Key Laboratory of Mathematical Science, Academy of Mathematics and Systems Science, Chinese Academy of Sciences, Beijing 100190, China}

\maketitle

\begin{abstract}
    Estimating multivariate traces \(\operatorname{Tr}(A_1A_2\cdots A_k)\) is a fundamental task arising across diverse fields. In this paper, We develop a quantum algorithm for this problem tailored to near-term quantum devices. 
    We access matrices by preparing their vectorized states. This access mode yields a tensor-network representation of the multivariate trace, which in turn guides the design of a parallel algorithm. The resulting circuit has depth dominated solely by that of the input oracle and independent of the number of matrices \(k\), at the cost of a qubit count scaling as \(\Theta(k)\)---a trade-off between circuit depth and circuit size. Moreover, the circuit exhibits a simple modular structure that allows explicit distribution over \(O(k)\) devices. We further establish efficient preparation circuits for a broad range of matrix classes, demonstrating the wide applicability of our approach. we present two applications: spectral sum estimation and a quantum implementation of the classical Lanczos method. Both reduce the target problem to sums of multivariate traces, thereby avoiding the construction of matrix polynomial transformations and achieving two levels of parallelism. For certain matrix classes, the resulting algorithms reduce circuit depth by a factor equal to the degree of the approximating polynomial. The shallow circuit depth and distributed implementation make our algorithms particularly suitable for near-term devices.
\end{abstract}

\section{Introduction}

We study the problem of approximating multivariate traces
\(\operatorname{Tr}(A_1A_2\cdots A_k)\) in this work. For density
matrices, this task is a fundamental building block in quantum information
science, with applications to estimating nonlinear functions of quantum
states \cite{PhysRevLett.88.217901}, such as integer R\'{e}nyi entropies
\cite{2017Entanglement,PhysRevLett.120.050406} and entanglement measures
\cite{Pawe2002Method,doi:10.1126/science.aau4963}. It is also a component of
quantum protocols such as quantum digital signatures \cite{gottesman2001quantumdigitalsignatures}.
For a general square matrix \(A\), the matrix moments
\(\operatorname{Tr}(A^k)=\sum_j\lambda_j^k\) reveal useful spectral
properties of \(A\). More generally, spectral sums
\(\operatorname{Tr}[f(A)]\) play a crucial role in many fields, ranging from
machine learning \cite{1756006.1953029,pmlr-v37-hana15,3524938.3525419} to
statistics \cite{2999325.2999396,golub1997generalized,dashti2011uncertainty}.

A naive strategy for estimating a multivariate trace would be to construct
the block encoding of the matrix product \(A_1A_2\cdots A_k\) from block
encodings of the individual matrices \(A_i\), and then apply a
trace-estimation subroutine to \(A_1A_2\cdots A_k\) to obtain an estimate of
\(\operatorname{Tr}(A_1A_2\cdots A_k)\). This results in a circuit depth of
\(\Omega(k)\) \cite{gilyen2019quantum}. In the current era of noisy
intermediate-scale quantum (NISQ) devices, circuit depths are severely
limited by short coherence times and gate errors. It is therefore desirable
to design quantum algorithms with shallow circuit depths.

In this paper, we provide a parallel quantum algorithm for multivariate
trace estimation whose circuit depth is independent of the number of
matrices \(k\). We access the matrices by preparing their vectorized state
(the state-preparation input model), and show that broad classes of matrices
admit efficient input circuits in this model. Matrix vectorization has a
natural tensor representation. Our construction is then guided by a
tensor-network representation of multivariate traces; transforming this
representation yields a method for computing traces of matrix products. Our
approach leverages graphical calculus---a tool with deep roots in physics and
quantum information science
\cite{abramsky2004categorical,selinger2007dagger,berezutskii2025tensor}---to
replace abstract symbolic manipulations with intuitive visual reasoning and
concrete computation.

The algorithm has a circuit depth dominated solely by the depth of the input
oracles, but requires a qubit count that scales as \(\Theta(k)\).
Fortunately, the circuit is highly structured, allowing a systematic
grouping of qubits into modules. We then provide an explicit distribution
scheme to implement the circuit over \(k\) quantum devices. Therefore, when
multiple quantum devices are available, our algorithm can be implemented
in shallow depth and in parallel.

Multivariate trace estimation can be applied to estimate spectral sums via
polynomial approximation. Previous work \cite{luongo} constructs matrix
polynomial transformations and employs a trace-estimation subroutine, and
therefore has a circuit depth that is linear in the degree of the
approximating polynomial. Our multivariate trace estimation algorithm can be
seen as a generalization of the trace-estimation subroutine and provides a
reduction in circuit depth by a factor equal to the degree of the polynomial
when applied to spectral sum estimation. Moreover, our framework can deal
with more general matrices than the positive semidefinite matrices
considered in \cite{luongo}.

As an application, we provide a parallel quantum implementation of the
classical Lanczos method. Quantum subspace diagonalization constitutes a
major class of approaches to eigenvalue problems of Hamiltonians. A
particularly successful subclass is based on Krylov subspaces, where the
basis is generated through repeated applications of the Hamiltonian to an
initial reference state. The key step is to construct the projected
Hamiltonian in the Krylov subspace. Existing quantum Krylov diagonalization
(QKD) algorithms construct non-orthogonal Krylov vectors and hence do not
exactly reproduce the orthogonality, stability, and convergence properties
of classical Lanczos-type methods.
Reference~\cite{rammal2026orthogonalquantumkrylovdiagonalisation}
reformulated the classical Lanczos recursion at the operator level,
generating an orthogonal Krylov basis based on block encoding and generalized
quantum signal processing (GQSP). We reduce the construction of the projected
Hamiltonian to multivariate trace evaluation, thus eliminating the need to
generate the Krylov basis and avoiding the use of GQSP. Consequently, our
implementation reduces the circuit depth by a factor equal to the dimension
of the Krylov subspace; its parallel nature also makes it more suitable for
near-term devices.

Both spectral sum approximation and quantum Lanczos implementation exhibit
two levels of parallelism. One comes from the distributed implementation of
the multivariate trace estimation algorithm. The other comes from
approximating the target quantity as a summation of multivariate traces that
can be evaluated simultaneously.

\section{Input Model\label{sec_input}}

In this work, we treat states and matrices on an equal footing by accessing both vectors and matrices through quantum state-preparation circuits. The column-major vectorization of a matrix $A = (A_1, \dots, A_{2^q}) \in \mathbb{C}^{2^q \times 2^q}$, denoted by $\operatorname{vec}(A)$, is the column vector obtained by stacking the columns of $A$ into a single column vector $(A_1^T,\dots,A_{2^q}^T)^T$.
A $2q$-qubit unitary $U_A$ is said to be the state-preparation input oracle for $A$ if
\begin{equation}
    U_A\ket{0}^{\otimes 2q} = \frac{1}{\|A\|_F}\operatorname{vec}(A) \eqcolon \vecket{A},
\end{equation}
where $\|\cdot\|_F$ denotes the Frobenius norm and $\vecket{A}$ is a unit vector. That is, the input oracle encodes the state $\vecket{A}$ in the first column of $U_A$. For a general matrix stored in classical memory, the corresponding vectorized state can be generated using the quantum state preparation algorithms of \cite{zhang2022quantum}. However, achieving the optimal circuit depth \(\Theta(q)\) requires an exponential number of ancillas in the worst case. In Appendix~\ref{app:class_mat}, we identify broad classes of matrices whose vectorized state can be prepared with polynomial circuit depth and ancilla overhead. In particular, for a matrix given as a linear combination of Pauli strings, \(A=\sum_{\ell=1}^{L}\alpha_{\ell}P_{\ell}\) with
\(L=\operatorname{poly}(q)\), the input oracle can be realized with circuit
depth \(d=O(\log q)\) using \(m=O(qL\log L)\) ancillas.

A widely used input model for non-unitary matrices is that of block encoding \cite{gilyen2019quantum}. In Appendix \ref{app:conversion}, we show that the two input models can be transformed into each other, although the conversion may not be efficient. Recently, Ref.~\cite{somma2026quantum} demonstrates exponential separations between these two input oracles. Specifically, the authors identify problems for which having access to the state-preparation input oracle can be more powerful than having access to the block-encoding oracle and vice versa. This suggests that the state-preparation input model may be a complementary technique for matrix input in quantum computing. For matrices given as linear combinations of Pauli strings, the construction protocol of Ref.~\cite{zhang2024circuit} for block-encoding exhibits a
trade-off between the number of ancilla qubits and the circuit depth:
using \(m\) ancilla qubits yields circuit depth
\(\tilde{O}(qL\log m/m)\), where
\(\Omega(\log_2 L)\le m\le O(qL)\). In particular, with \(O(qL)\) ancilla
qubits the circuit depth becomes \(O(\log(qL))\), indicating that the two
input models consume resources of the same order of magnitude.

\begin{figure}
    \centering
    \begin{subfigure}{0.3\columnwidth}
        \centering
        \begin{tikzpicture}[every node/.style={draw}]
            \node (A) at (0,0) {$A$};
            \draw (A.west)--(-0.75,0);
            \draw (A.east)--(0.75,0);
        \end{tikzpicture}
        \subcaption{}
    \end{subfigure}
    \begin{subfigure}{0.3\columnwidth}
        \centering
        \begin{tikzpicture}[every node/.style={draw}]
            \node (A) at (0,0) {$A$};
            \draw (A.west)--(-0.75,0);
            \draw (A.east)--(0.75,0)--(0.75,0.75)--(-0.75,0.75);
        \end{tikzpicture}
        \subcaption{}
    \end{subfigure}
    \begin{subfigure}{0.3\columnwidth}
        \centering
        \begin{tikzpicture}[every node/.style={draw}]
            \node (A) at (0,0) {$A$};
            \draw (A.west)--(-0.75,0)--(-0.75,0.75)--(0.75,0.75)--(0.75,0)--(A.east);
        \end{tikzpicture}
        \subcaption{}
    \end{subfigure}
    \caption{Tensor network representations of (a) $A$, (b) $\operatorname{vec}(A)$, and (c) $\operatorname{Tr}(A)$.}
    \label{input_model}
\end{figure}
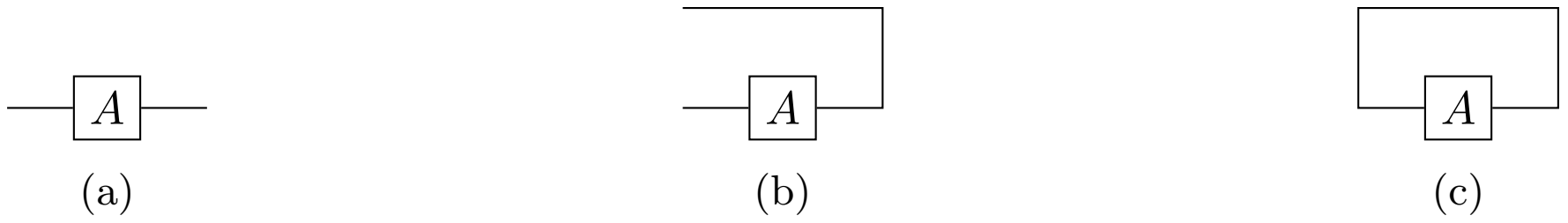

Tensor networks provide a graphical representation that facilitates the understanding and analysis of tensor operations. In this work, we focus on first-order tensors (vectors) and second-order tensors (matrices). A matrix $A = \sum\limits_{i,j}a_{ij}\ket{i}\bra{j}$ is represented as a box with two legs: the left leg corresponds to the row index, and the right leg corresponds to the column index. The column-major vectorization, $\operatorname{vec}(A) = \sum\limits_{i,j}a_{ij}\ket{j}\ket{i}$, is then represented by bending the right leg upward and to the left. Tensor contractions are represented by joining the corresponding legs; in particular, the matrix product $AB$ is represented by connecting the right leg of $A$ with the left leg of $B$, while the trace is obtained by connecting the row and column legs of the same matrix (Fig.~\ref{input_model}). We refer the reader to Appendix~\ref{app_tensor} for a more detailed introduction.

\section{Multivariate Trace Estimation}\label{multi_trace_est}

\begin{figure*}
    \centering
    \begin{tikzpicture}[every node/.style={draw}]
        \node (A1) at (0,0) {$A_1$};
        \node (A2) at (1,0) {$A_2$};
        \node[draw=none] (sym) at (2,0) {$\cdots$};
        \node (A2k-1) at (3.25,0) {$A_{2k-1}$};
        \node (A2k) at (4.5,0) {$A_{2k}$};
        \draw (A1.west)--++(-0.25,0)--++(0,-0.75)--($(A2k.east)+(0.25,0)+(0,-0.75)$)--++(0,0.75)--(A2k.east);
        \draw (A2k.west)--(A2k-1.east);
        \draw (A2k-1.west)--(sym.east);
        \draw (sym.west)--(A2.east);
        \draw (A2.west)--(A1.east);

        \node[draw=none] (op) at (5.4,-0.375) {$=$};

        \node (A1) at (6.25,0) {$A_1$};
        \node (A2k) at (7.25,-0.75) {$A_{2k}^T$};
        \node (A2k-1) at (8.5,-0.75) {$A_{2k-1}^T$};
        \node[draw=none] (sym) at (9.75,-0.75) {$\cdots$};
        \node (A2) at (10.75,-0.75) {$A_2^T$};
        \draw (A1.west)--++(-0.25,0)--++(0,-0.75)--(A2k.west);
        \draw (A2k.east)--(A2k-1.west);
        \draw (A2k-1.east)--(sym.west);
        \draw (sym.east)--(A2.west);
        \draw (A2.east)--++(0.25,0)--++(0,0.75)--(A1.east);

        \draw[dashed] ($(A1.east)!0.5!(A2k.west)+(0,0.75)$)--($(A1.east)!0.5!(A2k.west)+(0,-0.75)$);
    \end{tikzpicture}
    \caption{Tensor network representation of $\operatorname{Tr}(A_1A_2\cdots A_{2k})$, which is equal to the product of $\mathrm{vec}(A_1)^T$ and $\mathrm{vec}(A_{2k}^TA_{2k-1}^T \cdots A_2^T)$.}
    \label{tn_tr_2k}
\end{figure*}
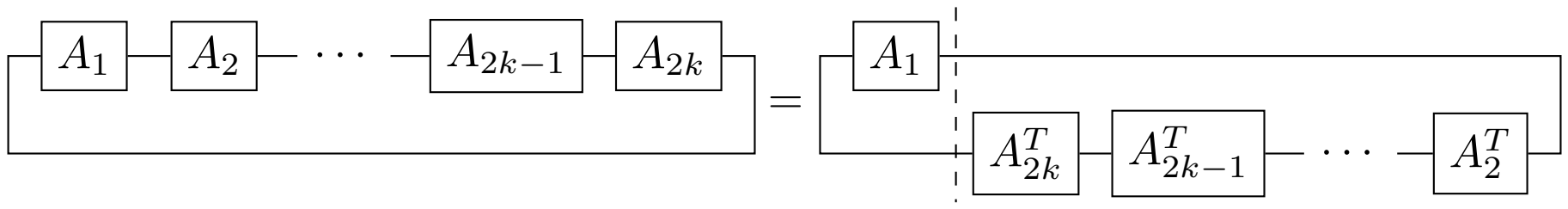

We consider the problem of estimating the trace of a product of an even number of matrices $\text{Tr}(A_1A_2\cdots A_{2k})$; the odd case is discussed in Appendix~\ref{app:odd}. The tensor network of a multivariate trace $\text{Tr}(A_1A_2\cdots A_{2k})$ is depicted in Fig.~\ref{tn_tr_2k}, which indicates that the multivariate trace is equal to the product of the row vector $\phi=\operatorname{vec}(A_1)^T$ and the column vector $\psi = \operatorname{vec}(A_{2k}^TA_{2k-1}^T \cdots A_2^T)$. It is known that $\operatorname{vec}(AB) = (I \otimes A) \operatorname{vec}(B)$. We generalize this formula to products of an arbitrary number of matrices, demonstrating that the vectorization of multiple matrix products can be expressed as matrix-vector multiplication.

\begin{figure}
    \centering
    \begin{tikzpicture}[every node/.style={draw}]
        \node (A1) at (0,0) {$A_1$};
        \node (A2) at (1,0) {$A_2$};
        \node (A3) at (2,0) {$A_3$};
        \node (A4) at (3,0) {$A_4$};
        \node (A5) at (2,-0.75) {$A_5^T$};
        \node (A6) at (1,-0.75) {$A_6^T$};
        \draw (A1.east)--(A2.west);
        \draw (A2.east)--(A3.west);
        \draw (A3.east)--(A4.west);
        \draw (A4.east)--++(0.25,0)--++(0,-0.75)--(A5.east);
        \draw (A5.west)--(A6.east);
        \draw (A6.west)--($(A1.west)+(-0.25,0)+(0,-0.75)$)--++(0,0.75)--(A1.west);

        \node[draw=none] (op) at (3.875,-0.375) {$=$};

        \node (A1) at (4.75,0.75) {$A_1$};
        \node (A2) at (6,0.75) {$A_2$};
        \node (A3) at (7,-0.75) {$A_3^T$};
        \node (A4) at (6,-0.75) {$A_4^T$};
        \node (A5) at (7,-1.5) {$A_5$};
        \node (A6) at (6,-0) {$A_6^T$};
        \draw (A1.west)--++(-0.25,0)--++(0,-0.75)--(A6.west);
        \draw (A1.east)--(A2.west);
        \draw (A2.east)--(7.5,0.75)--++(0,-1.5)--(A3.east);
        \draw (A3.west)--(A4.east);
        \draw (A4.west)--++(-0.15,0)--++(0,-0.75)--(A5.west);
        \draw (A5.east)--++(0.25,0)--++(0,1.5)--(A6.east);

        \draw[dashed] (5.375,1.25)--(5.375,-2);
        \draw[dashed] (6.5,1.25)--(6.5,-2);
    \end{tikzpicture}
    \caption{Transformation of the tensor network of $\operatorname{Tr}(A_1A_2A_3A_4A_5A_6)$ into a product of a row vector, a matrix, and a column vector.}
    \label{tn_trans_3}
\end{figure}
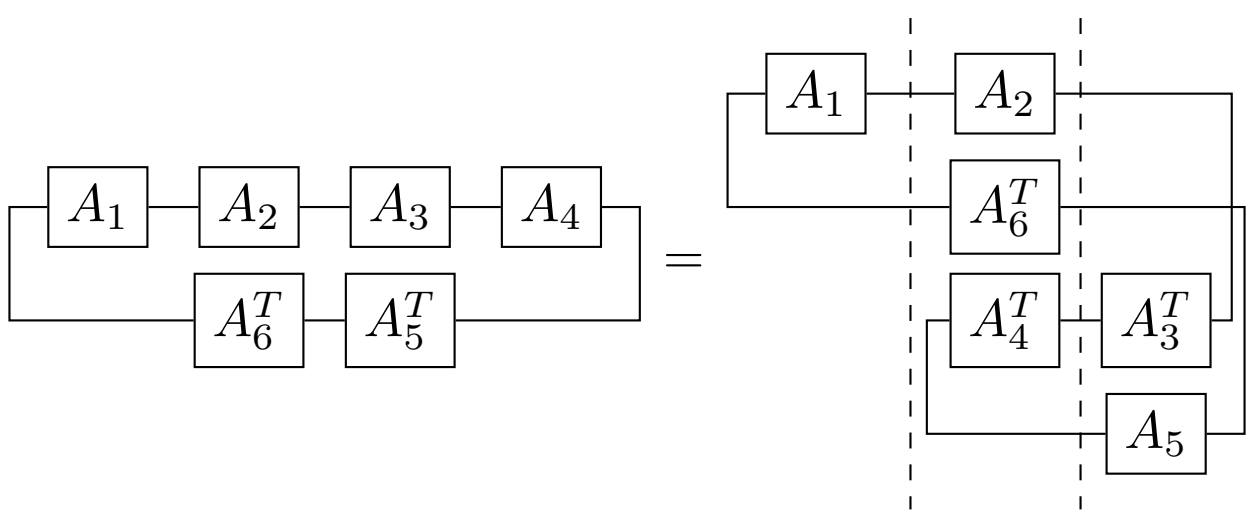

Taking $k=3$ as an example, Figure~\ref{tn_trans_3} shows that the tensor network of $\operatorname{Tr}(A_1A_2
A_3A_4A_5A_6)$ can be transformed and partitioned into three segments: a row vector $u=\operatorname{vec}(A_1)$, a matrix $M=(A_2\otimes A_6^T)\otimes\operatorname{vec}(A_4^T)^T$, and a column vector $v=\operatorname{vec}(A_3^T\otimes A_5)$. Therefore, we have
\begin{equation}
\begin{aligned}
    & \text{vec}(A_6^TA_5^TA_4^TA_3^TA_2^T) = A_2 \otimes A_6^T \otimes {\rm vec}(A_4^T)^T \cdot {\rm vec}(A_3^T \otimes A_5) = Mv, \\
    & {\rm Tr}(A_1A_2A_3A_4A_5A_6) = {\rm vec}(A_1)^T  \text{vec}(A_6^TA_5^TA_4^TA_3^TA_2^T) = u^TMv.
\end{aligned}
\end{equation}

Transformation of the multivariate trace into a product of three segments can be done similarly for $k>3$ by iteratively bending the legs. In general, we define matrices
\begin{equation}
    \begin{aligned}
        E_{i,k} &= A_{2i}^T \otimes A_{2(k+1-i)},\,i = 1,\dots,\frac{k-1}{2}, \\
        E_{\frac{k+1}{2},k} &= A_{k+1}^T, \\
        O_{i,k} &= A_{2i+1}^T \otimes A_{2(k-i)+1},\,i=1,\dots,\frac{k-1}{2},
    \end{aligned}
\end{equation}
for odd $k>1$, and
\begin{equation}
    \begin{aligned}
        E_{i,k} &= A_{2i}^T \otimes A_{2(k+1-i)},\,i = 1,\dots,\frac{k}{2}, \\
        O_{i,k} &= A_{2i+1}^T \otimes A_{2(k-i)+1},\,i=1,\dots,\frac{k}{2}-1, \\
        O_{\frac{k}{2},k} &= A_{k+1}^T.
    \end{aligned}
\end{equation}
for even $k>1$. The following lemma takes inspiration from \cite[Theorem 17]{mor2025multivariate}. However, we observe that the original statement and the proof have certain problems, which we discuss in Appendix \ref{app_err}. Rather than resorting to intricate algebraic derivations as in \cite{mor2025multivariate}, we are able to give a diagrammatic argument based on tensor networks. The proof is deferred to Appendix \ref{app_tr_eva}.

\begin{lemma}\label{tr_Mv}
The multivariate trace $\operatorname{Tr}(A_1A_2 \cdots A_{2k})$ can be expressed as matrix-vector multiplication $u^TMv$, where $u = \operatorname{vec}(A_1)$,
\begin{align}
    M &= E_{1,k}^T \otimes \operatorname{vec}(E_{2,k})^T \otimes \cdots \otimes \operatorname{vec}(E_{\lceil\frac{k}{2}\rceil,k})^T \label{M}, \\ 
    v &= \operatorname{vec}(O_{1,k}) \otimes \cdots \otimes \operatorname{vec}(O_{\lfloor\frac{k}{2}\rfloor,k}). \label{v}
\end{align}
\end{lemma}

Now assume that the matrices $A_1, A_2, \dots, A_{2k}$ can be accessed through the state-preparation oracles $U_1,U_2,\dots,U_{2k}$. To build the circuit for calculating $u^TMv$, we require the state-preparation unitaries $U_M$ and $U_v$ for $M$ and $v$, respectively. Observe that both $M$ and $v$ are formed as Kronecker products of matrices or vectorized matrices. The circuit implementations for $U_M$ and $U_v$ then follow from simple procedures that construct state-preparation oracles for Kronecker products of matrices, given oracles for the individual matrices (see Appendix \ref{app_tr_eva}). We introduce permutations $\sigma_i = (4i-2,4i-1)$, $\sigma=(2k-1,2k-3,\dots,5,3)(2k,2k-2,\dots,6,4)$, $\tau_{i} = (4i-3,4i-2)$, $\tau=(3,4)$, and $\tau'=(2k-1,2k)$, and denote by $S_i$, $S_\sigma$, $T_i$, $T_\tau$, $T_{\tau'}$ the corresponding permutations of the registers, respectively. The detailed proof of the following lemma is given in Appendix \ref{app_tr_eva}. Although the statement of the theorem appears rather complicated, its construction is essentially obtained by tensoring together some basic modules and has a simple and clear graphical intuition.

\begin{lemma}\label{input_M_v}
    Given the state-preparation oracles $U_2,U_3,\dots,U_{2k}$ for the matrices $A_2,A_3,\dots,A_{2k}$, define $V_1=U_2 \otimes (T_\tau U_{2k})$, $V_i = (T_iU_{2i}) \otimes U_{2(k+1-i)}$ for $i>1$, and $W_i = (T_i U_{2i+1}) \otimes U_{2(k-i)+1}$. The state-preparation unitaries for $M$ and $v$ as defined in Eqs.~\eqref{M} and~\eqref{v} can be constructed as follows:
    \begin{equation}
    \begin{aligned}
        U_M = & S_{\sigma} \cdot (S_1V_1) \otimes (S_2 V_2) \otimes \cdots \otimes (S_{\frac{k-1}{2}}V_{\frac{k-1}{2}}) \otimes (T_{\tau'} U_{k+1}), \\
        U_v = & (S_1W_1) \otimes (S_2 W_2) \otimes \cdots \otimes (S_{\frac{k-1}{2}}W_{\frac{k-1}{2}})
    \end{aligned}
    \end{equation}
    for odd $k>1$, and
    \begin{equation}
    \begin{aligned}
        U_M = & S_{\sigma} \cdot (S_1V_1) \otimes (S_2 V_2) \otimes \cdots \otimes (S_{\frac{k}{2}}V_{\frac{k}{2}}), \\
        U_v = & (S_1W_1) \otimes (S_2 W_2) \otimes \cdots \otimes (S_{\frac{k}{2}-1}W_{\frac{k}{2}-1}) \otimes (T_{\tau'} U_{k+1})
    \end{aligned}
    \end{equation}
    for even $k>1$.
\end{lemma}

The following lemma prepares a state that encodes $Mv = \mathrm{vec}(A_{2k}^TA_{2k-1}^T \cdots A_2^T)$ as a component, which is a direct application of the subroutine for matrix-vector multiplication under the state-preparation input model (see Appendix \ref{app_tensor}).

\begin{lemma}\label{mat_prod_vec}
    For matrices $A_2,A_3,\dots,A_{2k} \in \mathbb{C}^{2^q\times 2^q}$, $U_\psi = (U_v^T \otimes I_{2^{2q}}) U_M$ is a $2kq$-qubit quantum circuit that acts on $\ket{0}^{\otimes 2kq}$ as
    \begin{equation}
        \begin{aligned}
            U_\psi \ket{0}^{\otimes 2kq} = 
            & (U_v^T \otimes I_{2^{2q}}) U_M \ket{0}^{\otimes 2kq} \\ =
            & \ket{0}^{\otimes 2(k-1)q} \otimes \frac{\operatorname{vec}(A_{2k}^TA_{2k-1}^T\cdots A_2^T)}{\|A_2\|_F\|A_3\|_F \cdots \|A_{2k}\|_F} + \ket{\perp},
        \end{aligned}
    \end{equation}
    where $(\bra{0}^{\otimes 2(k-1)q}\otimes I_{2^{2q}})\ket{\perp} = 0$.
\end{lemma}

Post-selecting the first $2(k-1)q$ qubits to be in $\ket{0}$ yields the state $\frac{\operatorname{vec}(A_{2k}^TA_{2k-1}^T\cdots A_2^T)}{\|A_2A_3\cdots A_{2k}\|_F}$ on the remaining $2q$ qubits. This implies that $(\bra{0}^{\otimes 2(k-1)q} \otimes I_{2^{2q}})U_\psi(\ket{0}^{2(k-1)q}\otimes I_{2^{2q}})$ is the state-preparation operator (may not be unitary) for the matrix $A_{2k}^TA_{2k-1}^T \cdots A_2^T$.

Note that $\bar{U}_1\ket{0}^{\otimes 2q} = \frac{{\rm vec(\bar{A}_1)}}{\|A_1\|_F}$, we have 
\begin{equation}
\bra{0}^{\otimes 2q}\bar{U}_1^\dagger = \frac{{\rm vec}(A_1)^T}{\|A_1\|_F} = \frac{u^T}{\|A_1\|_F}.
\end{equation}
Let $U_u = U_1$. Since $U_\psi$ encodes $\frac{Mv}{\|A_2\|_F\|A_3\|_F \cdots \|A_{2k}\|_F}$ as a component in its first column, we take $U_\phi = I_{2^{2(k-1)q}} \otimes \bar{U}_u$ and obtain
\begin{equation}
    \begin{aligned}
    U_\phi^\dagger U_\psi\ket{0}^{\otimes 2kq} 
    &= I_{2^{2(k-1)q}} \otimes \bar{U}_u^\dagger \left(\ket{0}^{\otimes 2(k-1)q} \otimes \frac{Mv}{\|A_2\|_F\|A_3\|_F \cdots \|A_{2k}\|_F} + \ket{\perp} \right) \\
    &= \frac{u^TMv}{\|A_1\|_F\|A_2\|_F \cdots \|A_{2k}\|_F}\ket{0}^{\otimes 2kq} + \ket{\perp'} \\
    &= \frac{\operatorname{Tr}(A_1A_2\cdots A_{2k})}{\|A_1\|_F\|A_2\|_F \cdots \|A_{2k}\|_F}\ket{0}^{\otimes 2kq} + \ket{\perp'},
    \end{aligned}
\end{equation}
where $\ket{\perp'}$ is orthogonal to $\ket{0}^{\otimes 2kq}$. To extract the complex amplitude $\langle0^{\otimes 2kq}|U_\phi^\dagger U_\psi|0^{\otimes 2kq}\rangle$, we run the Hadamard test for the unitary 
\begin{equation}
U_\phi^\dagger U_\psi = (I_{2^{2(k-1)q}} \otimes U_u^T)(U_v^T \otimes I_{2^{2q}})U_M = (U_v^T \otimes U_u^T)U_M.
\end{equation}

\begin{figure}
    \centering
    \includegraphics[width=0.5\textwidth]{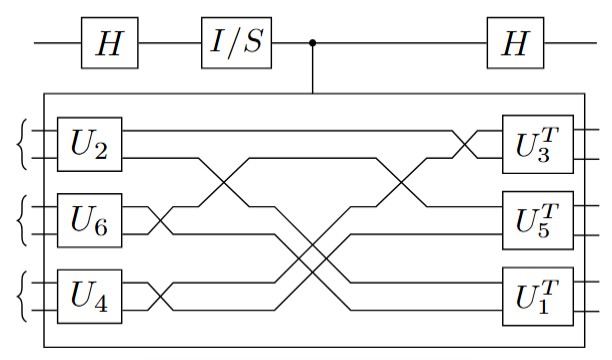}
    \caption{Multivariate trace estimation circuit for the case $k=3$: a Hadamard test circuit applied to the unitary $U_\phi^\dagger U_\psi$. Choosing $W=I$ or $W=S$ ($S$ is the phase gate) allows us to estimate the real or the imaginary part of $\langle0|^{\otimes 2kq}U_\phi^\dagger U_\psi|0\rangle^{\otimes 2kq}$. The permutation of wires corresponds to the permutations defined in Lemma \ref{input_M_v}, which can be implemented by SWAP gates.}
    \label{HT}
\end{figure}

\begin{theorem}\label{trace_est}
    The multivariate trace $\operatorname{Tr}(A_1A_2\cdots A_{2k})$ for $A_i \in \mathbb{C}^{2^q \times 2^q}$ can be estimated to within additive error $\epsilon$ with success probability at least $1-\eta$ by running the Hadamard test circuit for unitary $U_\phi^\dagger U_\psi$ $O(F^2\epsilon^{-2}\log(1/\eta))$ times, where $F=\prod_{i=1}^{2k}\|A_i\|_F$. 
\end{theorem}
\begin{proof}
    The rescaled multivariate trace $\frac{\mathrm{Tr}(A_1A_2 \cdots A_{2k})}{\|A_1\|_F\|A_2\|_F \cdots \|A_{2k}\|_F}$ can be estimated by running the Hadamard test circuit for the unitary $U_{\phi}^{\dagger}U_\psi$. It is well known that the Hadamard test requires repetitions of circuits $O(\epsilon^{-2}\log(1/\eta))$ times to achieve an additive error of $\epsilon$ with a success probability $1-\eta$ \cite{10.1145/1132516.1132579}. Thus, the sampling complexity for approximating $\operatorname{Tr}(A_1A_2\cdots A_{2k})$ is $O(\prod_{i=1}^{2k}\|A_i\|_F^2\epsilon^{-2}\log(1/\eta))$.
\end{proof}

The circuit for the case of $k=3$ is shown in Fig.~\ref{HT}. It has a highly parallel structure, which comes at the cost of an additional factor $\prod_{i=1}^{2k}\|A_i\|_F^2$ in complexity. This establishes a trade-off between circuit depth and sampling complexity, compared with the naive strategy. In Section~\ref{para_OQKD}, we present applications in which the norm-factor issue can be alleviated or bypassed.


\subsection{Distributed Implementation}

A remarkable feature of the circuit for multivariate trace estimation is that the circuit depth is independent of the number of matrices $2k$ and is simply dominated by the depth of the input oracles. A consequence of this high degree of parallelization is that the qubit count scales as $O(kq)$, i.e., linearly in \(k\) for fixed \(q\), which is undesirable for large $k$ or for near-term quantum devices with limited qubit counts. In this section, we show how the circuit implementation can be distributed over multiple quantum devices.

\begin{figure}
    \centering
    \includegraphics[width=0.8\linewidth]{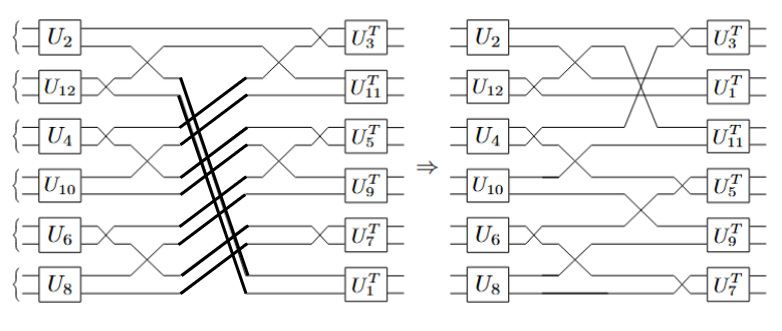}
    \caption{Eliminating the operation $S_\sigma$ (the bold wires). The curly braces highlight the grouping of qubits into modules.}
    \label{eg_6}
\end{figure}

We first note that the circuit can be further simplified. As illustrated in Fig.~\ref{eg_6} for the case of $k=6$, we permute the operators according to the permutation $\sigma$ so that we can remove $S_\sigma$ from the circuit. We then group qubits into modules as shown in Fig.~\ref{eg_6} and treat the module as a single quantum device. The only inter-device operations are wire permutations that can be implemented by SWAP gates.

\begin{figure}
    \centering
    \begin{subfigure}{0.45\textwidth}
        \centering
        \includegraphics[width=0.35\textwidth]{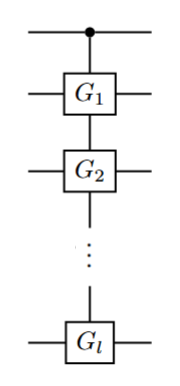}
        \subcaption{A layer of a controlled-$U$ operation: gates with a shared control.}
        \label{layer}
    \end{subfigure}
    \begin{subfigure}{0.45\textwidth}
        \centering
        \includegraphics[width=0.38\textwidth]{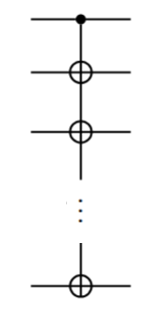}
        \subcaption{Quantum fan-out operation.}
        \label{fanout}
    \end{subfigure}
    \caption{}
\end{figure}

Now assume that \(U_\phi^\dagger U_\psi\) is compiled into one-qubit and CNOT gates, with each nonlocal SWAP gate decomposed into three CNOT gates. All these gates are controlled by a common qubit in the Hadamard-test circuit. As shown in Ref.~\cite{gokhale2021quantum}, controlled single-qubit gates and CNOT gates sharing a common control qubit can be executed simultaneously in constant depth (Fig.~\ref{layer}), allowing a controlled unitary to be implemented with the same circuit depth as its uncontrolled counterpart. 

\begin{figure}
    \centering
    \begin{subfigure}{0.45\textwidth}
        \centering
        \includegraphics[width=\textwidth]{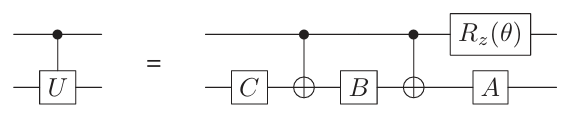}
        \caption{Circuit decomposition of a controlled single-qubit gate.}
        \label{cU}
    \end{subfigure}
    \begin{subfigure}{0.45\textwidth}
        \centering
        \includegraphics[width=\textwidth]{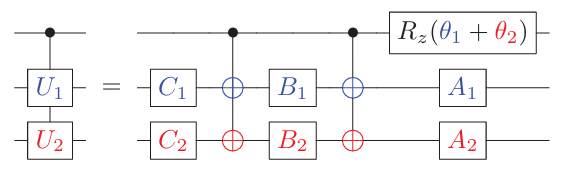}
        \caption{Circuit synthesis for shared-control single-qubit gates.}
        \label{cUU}
    \end{subfigure}
    \caption{}
\end{figure}

\begin{figure}
    \centering
    \begin{subfigure}{0.45\textwidth}
        \centering
        \includegraphics[width=\textwidth]{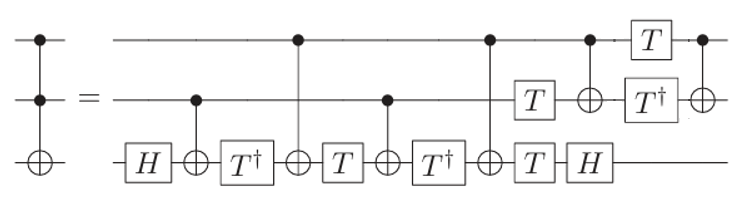}
        \caption{Circuit decomposition of a Toffoli gate.}
        \label{toffoli}
    \end{subfigure}
    \begin{subfigure}{0.45\textwidth}
        \centering
        \includegraphics[width=\textwidth]{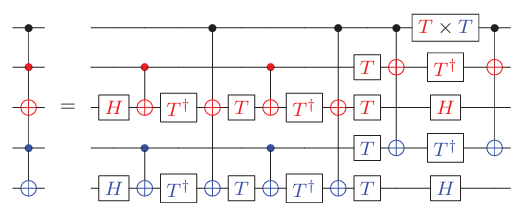}
        \caption{Circuit synthesis for shared-control CNOT gates.}
        \label{cTT}
    \end{subfigure}
    \caption{}
\end{figure}

The synthesis procedure consists of two basic cases. The first case concerns shared-control single-qubit gates. A controlled single-qubit gate is typically decomposed as in Fig.~\ref{cU}, and the simultaneous execution of two single-qubit gates controlled by a common qubit is illustrated in Fig.~\ref{cUU}. The second case concerns shared-control CNOT gates. For a Toffoli gate decomposed as in Fig.~\ref{toffoli}, the simultaneous execution of two CNOT gates controlled by a common qubit is shown in Fig.~\ref{cTT}. The generalization to multiple gates sharing a common control qubit is straightforward. Observe that the circuits are decomposed into single-qubit gates, CNOT
gates, and quantum fan-out operations (Fig.~\ref{fanout}).

Applying this synthesis approach to the controlled-\(U_\phi^\dagger U_\psi\)
circuit yields a circuit consisting of local gates within each module and
non-local gates in the form of CNOT gates and quantum fan-out operations.
A distributed implementation of the CNOT gate has been established via gate
teleportation, requiring one entangled pair and two classical bits of
communication \cite{yimsiriwattana2004generalizedghzstatesdistributed}.
The quantum fan-out can be viewed as a controlled-\(U\) operation, where
\(U\) is a tensor product of gates. This type of controlled operation can be
readily distributed and executed in parallel
\cite{yimsiriwattana2004generalizedghzstatesdistributed}.

\begin{theorem}
    Assume that the $2^q \times 2^q$ matrices \(A_1,A_2,\dots,A_{2k}\) are accessed through state-preparation oracles \(U_1,U_2,\dots,U_{2k}\), each of circuit depth \(O(d)\) and using \(O(m)\) ancilla qubits. The circuit for multivariate trace estimation can be synthesized into \(\Theta(k)\) modules of size $\Theta(q)$, with CNOT gates and quantum fan-outs as the only inter-module operations.
    
    Assume that the synthesized circuit contains $N_{\mathrm{CX}}$ = O(kq) CNOT gates and \(N_f=O(d)\) quantum fan-out operations. Then the multivariate trace estimation circuit can be distributed over $\Theta(k)$ quantum devices and runs in time \(O(d+N_f\log k)\), consuming \(O(N_{\mathrm{CX}}+kN_f)\) entangled pairs and \(O(N_{\mathrm{CX}}+kN_f)\) bits of classical communication. Each module uses \(O(m)\) ancillas.
\end{theorem}

\begin{proof}
The number of
non-local CNOT gates is on the order of the product of the
number of modules and the module size, as illustrated in Figs.~\ref{eg_6} and~\ref{cTT}, while the number of
quantum fan-out operations is of the same order as the depth of the unitary
being synthesized in this systhesis approach. Hence, $N_{\mathrm{CX}} = O(kq)$ and $N_f = O(d)$.

A quantum fan-out across \(k\) modules can be distributed as follows (see
Appendix~\ref{app:fanout} for details). First, generate a distributed \(k\)-fold generalized GHZ
state, which can be implemented in \(O(\log k)\) steps at a cost of
\(k-1\) entangled pairs and \(4(k-1)\) bits of classical communication.
Then the cat-entangler distributes the control among the \(k\) modules,
and the quantum fan-out, viewed as a composition of multiple controlled
gates, can be implemented in parallel. Finally, the cat-disentangler
restores the control qubit. Both the cat-entangler and the cat-disentangler
run in constant time and require \(2(k-1)\) bits of classical communication.
The resource cost stated in the theorem follows.
\end{proof}



\subsection{Spectral Sums}

Multivariate trace estimation can be used for spectral sum estimation. A spectral sum is defined as follows.
\begin{definition}
Let the eigenvalues of \(A\) be \(\{\lambda_j\}\), counted with their
algebraic multiplicities. Let $f \colon \mathbb{C} \to \mathbb{C}$ be a function. The spectral sum of $A$ for the function $f$ is defined as $S_f(A) = \mathrm{Tr}[f(A)] = \sum_j f(\lambda_j)$.
\end{definition}

One of the most famous examples of spectral sums is log-determinant $\log\det(A) = \sum_j \log\lambda_j$, which arises in machine learning \cite{1756006.1953029,pmlr-v119-kerenidis20a} and other computational sciences \cite{rue2005gaussian}.

Ref.~\cite{mor2025multivariate} suggested two methods for this task, which can be implemented straightforwardly using our parallel algorithm. Suppose that $p(x) = \sum_{i=0}^n c_ix^i$ is a polynomial of degree \(n\). The first is to use matrix moments. Since
\begin{equation}
    \mathrm{Tr}[p(A)] = \sum_{i=0}^n c_i\mathrm{Tr}(A^i),
\end{equation}
we can use our algorithm to approximate each \(\operatorname{Tr}(A^i)\), \(i=0,\dots,n\), as a multivariate trace. Moreover, these quantities can be computed simultaneously and then summed classically. The second approach reduces $\mathrm{Tr}[p(A)]$ to a single multivariate trace. We first factor the polynomial:
\begin{equation}
    p(x) = c_n(x-r_1)(x-r_2)\cdots(x-r_n),
\end{equation}
where $r_1,r_2,\dots,r_n$ are the roots of $p(x)$, counted with multiplicity. Then
\begin{equation}
    p(A) = c_n(A-r_1I)(A-r_2I)\cdots(A-r_nI).
\end{equation}
Thus, $\mathrm{Tr}[p(A)] = c_n\mathrm{Tr}[(A-r_1I)(A-r_2I)\cdots(A-r_nI)]$. Given a polynomial approximation \(p\) to the function \(f\),
the spectral sum \(\operatorname{Tr}[f(A)]\) can be approximated by
\(\operatorname{Tr}[p(A)]\) using either of the above two methods. The complexity would depend on the polynomial $p$ that approximates the function $f$. We state the theorem for the moment
method; the factorization approach can be analyzed similarly.

\begin{theorem}
Assume that the \(N\times N\) matrix \(A\) is accessed through the
state-preparation input oracle. Let $a \ge \|A\|_F$. For a function \(f\), assume access to an even
polynomial \(p(x)=\sum_{i=0}^{n/2}c_i x^{2i}\) of degree \(n\) that approximates \(f\) on a domain containing
the eigenvalues of \(A\) with error at most \(\epsilon/(2N)\).
Then the spectral sum \(\operatorname{Tr}[f(A)]\) can be estimated to within
error \(\epsilon\) with success probability at least \(1-\eta\) by
estimating the \(n\) multivariate traces
\(\operatorname{Tr}(A^i)\), \(i=2,4,\dots,n\), each with the multivariate
trace estimation circuit run
\(O(C_p^2\epsilon^{-2}\log(n/\eta))\) times.
The total number of circuit executions is
\[
    O\!\left(C_p^2 n\epsilon^{-2}\log(n/\eta)\right),
\]
where \(C_p=\sum_{i=0}^{n/2}|c_i|a^{2i}\).
\end{theorem}

\begin{proof}
Let \(\hat A=A/a\) with \(a\ge\|A\|_F\), so that
\(\|\hat A\|_F\le1\). By assumption,
\begin{equation}
    \max_{x\in\sigma(A)}|f(x)-p(x)|
    \le
    \epsilon_1
    =
    \epsilon/(2N).
\end{equation}
Assume that the multivariate traces are estimated with error at most
\(\epsilon_2\):
\begin{equation}
    \left|
    \operatorname{Tr}(\hat A^i)
    -
    \widetilde{\operatorname{Tr}}(\hat A^i)
    \right|
    \le
    \epsilon_2,
    \qquad
    i=2,4,\dots,n.
\end{equation}
Then
\begin{equation}
\begin{aligned}
    \left|
    \operatorname{Tr}[f(A)]
    -
    \widetilde{\operatorname{Tr}}[p(A)]
    \right|
    &\le
    \left|
    \operatorname{Tr}[f(A)-p(A)]
    \right|
    +
    \left|
    \operatorname{Tr}[p(A)]
    -
    \widetilde{\operatorname{Tr}}[p(A)]
    \right| \\
    &\le
    N\epsilon_1+C_p\epsilon_2.
\end{aligned}
\end{equation}
To make the right-hand side at most \(\epsilon\), it suffices to take
\(\epsilon_2=\epsilon/(2C_p)\). The theorem then follows from the
complexity of multivariate trace estimation in Theorem~\ref{trace_est}.
The value of the failure-probability parameter \(\eta\) is obtained from a
standard union bound argument.
\end{proof}

In practice, the moment method is expected to be easier to implement, since it requires only an input oracle for \(A\), whereas the factorization approach requires input oracles for shifted matrices. Moreover, the norm factor
\(\prod_{i=0}^{n}\|A-r_iI\|_F^2\) in the factorization method may be
larger, and the root-finding step may suffer from numerical instability.

Recently, Ref.~\cite{luongo} propose quantum algorithms for estimating spectral sums. Specifically, they present quantum algorithms to estimate the log-determinant, the von Neumann entropy of a graph, the trace of inverse, and the Schatten $p$-norm, and consider applications in spectral graph theory. The key component is a quantum subroutine for estimating the trace of a matrix accessed through the block-encoding input model. Their approach for approximating $\mathrm{Tr}[f(A)]$ is to construct a polynomial approximation $p(A)$ to $f(A)$ using Quantum Singular Value Transformation (QSVT) techniques and then apply the trace estimation subroutine to $p(A)$.

\begin{figure}
\centering
    \begin{tikzpicture}[
    tensor/.style={draw}
    ]
        \node[tensor] (A) at (0,0) {$A$};
        \draw (A.west) --++ (-0.5,0) --++ (0,0.75) -- ($(A.east)+(0.5,0.75)$) --++ (0,-0.75) -- (A.east);

        \node[draw=none] (op) at (1.1,0.375) {$=$};

        \node[tensor] (B) at (2.5,0) {$A$};
        \draw (B.west) -- (1.5,0) --++ (0,0.75) -- ($(B.east)+(0.5,0.75)$) --++ (0,-0.75) -- (B.east);

        \draw[dashed] (1.85,1) -- (1.85,-0.5);
\end{tikzpicture}
\caption{$\mathrm{Tr}(A)$ is the inner product of vector $\mathrm{vec}(A)$ and vector $\mathrm{vec}(I)$.}\label{trace_subroutine}
\end{figure}
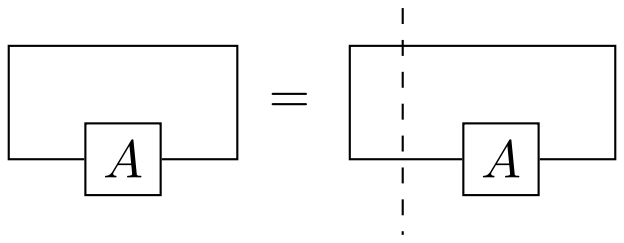

Here we reinterpret the idea of \cite{luongo} for trace estimation, which is illustrated in Fig.~\ref{trace_subroutine}. The quantity \(\operatorname{Tr}(A)\) can be written as the bilinear inner product
\(\operatorname{vec}(A)^T\operatorname{vec}(I)\) of the vectors \(\operatorname{vec}(A)\) and \(\operatorname{vec}(I)\). The quantum subroutine of \cite{luongo} essentially constructs the unitaries that prepare the normalized states corresponding to the vectors $\mathrm{vec}(A)$ and $\mathrm{vec}(I)$, and then applies the Hadamard test circuit to estimate the inner product. Constructing the unitary that prepares $\mathrm{vec}(A)$, given block encoding of $A$, is precisely the conversion of the block-encoding input model to the state-preparation input model (see Appendix~\ref{app:conversion} for a detailed discussion). Given this insight, our algorithm generalizes the trace estimation subroutine to multivariate trace. 

By comparison, our approach offers two advantages. First, our algorithm does
not rely on the Quantum Singular Value Transformation (QSVT) framework and
can therefore be applied to more general classes of matrices than the
positive semidefinite matrices considered in \cite{luongo}. Second, the
circuit depth of our algorithm is independent of the degree of the
approximating polynomial and is dominated by the circuit depth of the input
oracle. As mentioned in Section~\ref{input_model}, for matrices given as
linear combinations of Pauli strings, the two input models require nearly
the same resources. However, the circuit depth in \cite{luongo} is the
product of the query complexity and the circuit depth of the block encoding;
our approach therefore reduces the circuit depth by a factor of the
polynomial degree. The cost is a potentially higher sampling complexity,
i.e., a larger number of circuit executions.

We state the theorem only for functions that can be approximated by even
polynomials. For functions approximated by polynomials of indefinite
parity, this approach requires a multivariate trace algorithm for an
arbitrary number of matrices. In Appendix~\ref{sec:odd}, we discuss how to
treat multivariate traces of an odd number of matrices; for spectral sums,
this requires an efficient decomposition of one of the matrices.
Nonetheless, restricting to functions approximated by even polynomials
already captures several applications, including log-determinants, Schatten
\(p\)-norms for even \(p\), and practical applications such as counting
spanning trees and computing effective resistance \cite{luongo}.

\section{A Parallel Quantum Lanczos Algorithm}\label{para_OQKD}

Quantum subspace diagonalization is a major class of quantum algorithms for
eigenvalue estimation. In these algorithms, a set of basis states, referred
to as the Krylov basis, is generated and used to construct the projected
Hamiltonian on the subspace spanned by this basis. Solving the resulting
projected eigenvalue problem then yields approximations to the low-energy
spectrum of the original Hamiltonian. Compared with variational quantum
algorithms, subspace methods offer more systematic convergence and avoid
difficult nonlinear optimization landscapes. Compared with quantum phase
estimation, which requires fault-tolerant quantum computers, subspace
methods require shallower circuits and are therefore more suitable for
near-term devices. However, existing subspace diagonalization algorithms
generate non-orthogonal Krylov vectors and must therefore contend with the
ill-conditioning of the resulting generalized eigenvalue problem
\cite{Kirby_2023,yoshioka2025krylov,motta2020determining}. Consequently,
they do not fully reproduce the stability and convergence properties of
classical Lanczos-type methods \cite{doi:10.1137/0717059}.


The recently proposed Orthogonal Quantum Krylov Diagonalization (OQKD)
framework \cite{rammal2026orthogonalquantumkrylovdiagonalisation}
reformulates the classical Lanczos recursion in a quantum setting.
It generates an orthogonal Krylov basis by applying matrix polynomials to
the initial state, thereby reproducing the convergence properties of the
classical Lanczos algorithm \cite{doi:10.1137/0717059}. The circuit depth is
determined by the polynomial degree, which is upper-bounded by the dimension
\(D\) of the Krylov subspace. The value of \(D\) depends on the spectral
properties of the Hamiltonian and typically scales polynomially with the
number of qubits
\cite{rammal2026orthogonalquantumkrylovdiagonalisation}. In this section,
we show that the OQKD framework can be implemented using our parallel
multivariate trace estimation algorithm, leading to a reduction in circuit
depth by a factor of \(D\).

\subsection{Review of the Classical Lanczos Method and the OQKD Framework}
We briefly review the classical Lanczos method and the OQKD framework. Given a Hermitian matrix $H$ and a starting vector $\ket{\phi_0}$, the Lanczos algorithm successively generates an orthonormal basis $\{\ket{\psi_0},\cdots,\ket{\psi_{D-1}}\}$ (i.e., the Krylov basis) of the Krylov subspace $\mathcal{K}_D(H,\ket{\phi_0}) = \mathrm{span}\{ \ket{\phi_0},H\ket{\phi_0},\cdots,H^{D-1}\ket{\phi_0} \}$. Let \(P = [\ket{\psi_0}\ \ket{\psi_1}\ \cdots\ \ket{\psi_{D-1}}]\). In this basis, the projected Hamiltonian $H_D = P^\dagger HP$ is represented by the tridiagonal matrix
\begin{equation}
    H_D = \begin{pmatrix} a_0 & b_1 & & & & \\ b_1 & a_1 & b_2 & & & \\ & b_2 & a_2 & \ddots & & \\ & & \ddots & \ddots & \ddots & \\ & &  & \ddots & a_{D-2} & b_{D-1} \\ & & & & b_{D-1} & a_{D-1} \end{pmatrix}.
\end{equation}
It follows that 
\begin{equation}
    H\ket{\psi_j} = b_j\ket{\psi_{j-1}}+a_j\ket{\psi_j}+b_{j+1}\ket{\psi_{j+1}}.
\end{equation}
The Krylov basis and the coefficients $a_j$ and $b_j$ can be computed recursively as follows:
\begin{equation}
\begin{aligned}
    a_j &= \bra{\psi_j}H\ket{\psi_j}, \\
    b_{j+1} &= \| H\ket{\psi_j}-a_j\ket{\psi_j}-b_j\ket{\psi_{j-1}} \| = \sqrt{\bra{\psi_j}H^2\ket{\psi_j}-a_j^2-b_j^2}, \\
    \ket{\psi_{j+1}} &= \frac{1}{b_{j+1}}(H\ket{\psi_j}-a_j\ket{\psi_j}-b_j\ket{\psi_{j-1}}),
\end{aligned}
\end{equation}
where we set \(b_0=0\) and \(\ket{\psi_{-1}}=0\). Solving the projected eigenvalue problem 
\begin{equation}\label{eig_equ}
    H_Dv=\lambda v
\end{equation}
yields approximate eigenvalues \(\tilde{\lambda}\) of the original Hamiltonian, with the Ritz-vector \(P\tilde{v}\) being the corresponding approximate eigenstate.

Building on the Lanczos algorithm, Ref.~\cite{rammal2026orthogonalquantumkrylovdiagonalisation} observed that each Krylov basis state can be generated by applying a polynomial in the Hamiltonian to the initial state,
\begin{equation}
    \ket{\psi_n} = P_n(H)\ket{\phi_0}.
\end{equation}
The polynomials $P_n(\cdot)$, referred to as Lanczos polynomials, satisfy the Lanczos recurrence
\begin{equation}\label{poly_recur}
  b_{n+1}P_{n+1}(x) = (x-a_n)P_n(x)-b_nP_{n-1}(x),
\end{equation}
with 
\begin{equation}
    P_0(x)=1,\qquad P_1(x)=\frac{x-a_0}{b_1}.
\end{equation}
Reference~\cite{rammal2026orthogonalquantumkrylovdiagonalisation} then expands $P_n(\cdot)$ in the Chebyshev basis and employs generalized quantum signal processing (GQSP) to implement the polynomial transformation of the Hamiltonian under the block-encoding input model. The rapid growth of the polynomial coefficients causes the success probability of GQSP to decay approximately exponentially with the polynomial degree. To alleviate this
issue, the authors propose a restarted OQKD protocol.

\subsection{Parallel Implementation}
In this section, we show that the classical Lanczos method admits a natural and efficient quantum implementation within our framework. By combining our parallel multivariate trace estimation algorithm with the Lanczos polynomial transformation of the OQKD framework, we obtain a parallel quantum Lanczos method that can be distributed across multiple devices. We also provide a rigorous error analysis of our algorithm.

The quantum part of the OQKD algorithm consists of preparing the Krylov basis state \(\ket{\psi_n}\) and estimating \(\bra{\psi_n}H\ket{\psi_n}\) and \(\bra{\psi_n}H^2\ket{\psi_n}\) via measurements. We show that this task can be reduced to multivariate trace estimation without explicitly generating the Krylov basis.

Let $P_n(x) = \sum_i c_{n,i}x^i$. By Eq.~\eqref{poly_recur}, the polynomial coefficients obey the recurrence relation
\begin{equation}\label{coe_recur}
    c_{n+1,i} = \frac{c_{n,i-1}-a_nc_{n,i}-b_nc_{n-1,i}}{b_{n+1}}.
\end{equation}
Since the Lanczos coefficients are real, \(P_n(H)\) is self-adjoint, and hence
\begin{equation}\label{sum}
\begin{aligned}
    \bra{\psi_n}H\ket{\psi_n} &= \bra{\phi_0}P_n(H)HP_n(H)\ket{\phi_0} = \sum_{i,j} c_{n,i}c_{n,j} \bra{\phi_0}H^{i+j+1}\ket{\phi_0}, \\
    \bra{\psi_n}H^2\ket{\psi_n} &= \bra{\phi_0}P_n(H)H^2P_n(H)\ket{\phi_0} = \sum_{i,j} c_{n,i}c_{n,j} \bra{\phi_0}H^{i+j+2}\ket{\phi_0}.
\end{aligned}
\end{equation}
To evaluate the quantities $\bra{\psi_n}H\ket{\psi_n}$ and $\bra{\psi_n}H^2\ket{\psi_n}$, we estimate the matrix moments
\begin{equation}\label{tr_momment}
\bra{\phi_0}H^k\ket{\phi_0}
= \operatorname{Tr}\!\left(H^k\ket{\phi_0}\bra{\phi_0}\right)
\end{equation}
for \(k=1,2,\dots,2D\) simultaneously using our multivariate trace algorithm and then sum the results classically according to Eq.~\eqref{sum}.

To reduce the norm factor in the complexity, we normalize \(H\) as
\(K=H/h\) with \(h\ge\|H\|_F\), so that \(\|K\|_F\le1\). Let
\(X=(\ket{\phi_0},K\ket{\phi_0},\ldots,K^{D-1}\ket{\phi_0})\), and
redefine \(P\) as the matrix with orthonormal columns obtained by applying
Gram--Schmidt orthonormalization to the columns of \(X\), so that
\(P=XC\). Then the columns \(c_j\) of \(C\) are the coefficient vectors of
the Lanczos polynomials \(p_j\). Assume that
\begin{equation}
    \max_{0\le j<D}\|c_j\|_1\le\Gamma.
\end{equation}
The polynomial coefficients grow rapidly as the degree increases, since
each iteration involves division by a factor \(b_{n}<1\)
(Eq.~\eqref{coe_recur}). This issue can be alleviated by the restart
protocol used in
Ref.~\cite{rammal2026orthogonalquantumkrylovdiagonalisation}; hence, in
what follows, we assume that the \(1\)-norm bound \(\Gamma\) is reasonably
scaled.

\begin{theorem}\label{parallel_eig}
Let \(H\) be an \(N\times N\) Hermitian matrix with eigenvalues
\(\lambda_1<\lambda_2<\cdots<\lambda_N\), where \(N=2^q\).
Let \(W=\lambda_N-\lambda_1\) be the spectral width,
\(\Delta=\lambda_2-\lambda_1\) the spectral gap, and
\(S_0:=\lambda_N-\lambda_2\).
Let \(\theta_0\) be the angle between the initial state and the lowest eigenvector. Assume that $K=H/h$ can be accessed through the state-preparation input oracle.

Then the classical Lanczos method can be implemented quantumly and outputs an $\epsilon$-estimate of $\lambda_1$ with success probability at least $1-\eta$ by estimating $\Theta(D)$ matrix moments, each with the multivariate trace estimation circuit run $O(\Gamma^4\max\{ h^2\epsilon^{-2},D^2 \}\log(D/\eta))$ times. The total number of circuit execution is
\begin{equation}
    O(D\Gamma^4\max\{ h^2\epsilon^{-2},D^2 \}\log(D/\eta)).
\end{equation}
Here
\begin{equation}
    D = 1+
    \left\lceil
    \frac{1}{2a}
    \log\left(
    \frac{8W\tan^2\theta_0}{\epsilon}
    \right)
    \right\rceil,
\end{equation}
where $h \ge \|H\|_F$, \(a=\operatorname{arcosh}\gamma\), and
\(\gamma=1+2\Delta/S_0\).
\end{theorem}

The proof is given in Appendix~\ref{app:proof_eig}. For the regime \(\Delta/S_0\ll 1\), we have
\(a=\operatorname{arcosh}(1+2\Delta/S_0)\sim 2\sqrt{\Delta/S_0}\), and hence
\begin{equation}
    D \approx 1+
    \left\lceil
    \frac{1}{4}\sqrt{\frac{S_0}{\Delta}}
    \log\left(
    \frac{8W\tan^2\theta_0}{\epsilon}
    \right)
    \right\rceil .
\end{equation}
Our approach is most useful for Hermitian matrices whose Frobenius norm and
spectral-width-to-spectral-gap ratio are polynomially bounded.
We note that a similar error analysis can be performed for the smallest \(D\) eigenvalues using the convergence result for the classical Lanczos method
\cite{doi:10.1137/0717059}.

Another application is eigenvector approximation. Here we focus on extracting
information from the lowest eigenvector of a Hermitian matrix, namely, estimating the expectation value
of some observable with respect to the lowest eigenvector. This task can be accomplished through multivariate trace estimation without explicitly generating the eigenvector.

Let $\ket{\lambda_1}$ be the lowest eigenvector of $K$ and 
\begin{equation}
    \ket{\lambda_1^{\mathcal{K}}} = Pv_1 = XCv_1
\end{equation}
be the Ritz vector. Obtain $\tilde{K}_D$ and $\tilde{C}$ as estimations of $K_D$ and $C$ by the same procedure as in Theorem~\ref{parallel_eig}, and obtain $\tilde{{v}}_1$ from solving the eigenvalue equation of $\tilde{K}_D$. Assume an upper bound $V$ on the $1$-norm of the normalized lowest eigenvector $\tilde{v}_1$ of $\tilde{K}_D$. A universal choice is $V = \sqrt{D}$. 

The expectation values of observable $O$ with respect to $\ket{\lambda_1}$ and $\ket{\lambda_1^{\mathcal{K}}}$ are denoted, respectively, by
\begin{equation}
    \begin{aligned}
    & o = \bra{\lambda_1}O\ket{\lambda_1}, \\
    & o_D = \bra{\lambda_1^{\mathcal{K}}}O\ket{\lambda_1^{\mathcal{K}}} = v_1^\dagger C^TX^\dagger OXCv_1.
    \end{aligned}
\end{equation}
Let $\Omega = X^\dagger OX$ with $\Omega_{ij} = \bra{\phi_0}K^iOK^j\ket{\phi_0}$. We estimate the observable moments $\Omega_{ij}$ using our multivariate trace algorithm. Set $\tilde{d} = \tilde{C}\tilde{v}_1$, and output 
\begin{equation}
    \tilde{o} = \mathrm{Re}(\tilde{d}^\dagger \tilde{\Omega}\tilde{d})
\end{equation}
as our estimation of $o$.

\begin{theorem}\label{parallel_eigstate}
    Let $H$ be a Hermitian matrix with spectral width $W$ and spectral gap $\Delta$. $S_0 \coloneq W-\Delta$. Let observable $O$ be a Pauli string. Assume $K=H/h$ can be accessed through the state-preparation input model.
    
    The expectation value of \(O\) with respect to the lowest eigenvector \(\ket{\lambda_1}\) of \(H\) can be estimated to within error \(\epsilon\) with success probability at least \(1-\eta\) by estimating \(O(D)\) matrix moments, each requiring \(O\!\left(D\Gamma^4\max\{\Lambda^2\epsilon^{-2},D\}\log(D/\eta)\right)\) runs of the multivariate trace estimation circuit, and \(O(D^2)\) observable moments, each requiring \(O\!\left(\Gamma^4\Lambda^2\epsilon^{-2}\log(D/\eta)\right)\) runs of the same circuit, where \(\Lambda=\max\{V^2,h/\Delta\}\). The total number of circuit execution is
    \begin{equation}
        O(D^2\Gamma^4\max\{ \Lambda^2\epsilon^{-2},D \}\log(D/\eta)).
    \end{equation}
    Here
    \begin{equation}
        D = 1 + \left\lceil \frac{1}{a}\log\left( \frac{16 B\tan\theta_0}{\epsilon} \right) \right\rceil,
    \end{equation}
    where \(a=\operatorname{arcosh}\gamma\), \(\gamma=1+2\Delta/S_0\)$, B= \sqrt{1+\left(\frac{W}{\Delta}\right)^2}$.
\end{theorem}

The proof is given in Appendix~\ref{app:eigstate}. A theorem for the lowest
\(D\) eigenvectors can be stated similarly using the convergence result for
the classical Lanczos method \cite{doi:10.1137/0717059}. In the regime $\Delta/S_0 \ll 1$, we have $a = \mathrm{arcosh}(1+2\Delta/S_0) \sim 2\sqrt{\Delta/S_0}$, and hence
\begin{equation}
    D \approx 1+\frac{1}{2}\sqrt{\frac{S_0}{\Delta}}\log\left( \frac{16 B\tan\theta_0}{\epsilon} \right).
\end{equation}

Note that the quantum computing part need not proceed step by step along the Lanczos recurrence. Once $D$ is fixed, all required matrix moments and observable moments can be calculated in advance, so the quantum circuits need not wait for the recurrence coefficients or the Ritz vectors to be computed. These parts are deferred to the subsequent small-scale classical post-processing. Moreover, the task decomposes into independent quantum circuits; given enough devices, the multivariate traces required by the algorithm can be computed simultaneously.



The parallel quantum Lanczos algorithm can also serve as a tool for quantum
principal component analysis. Reference~\cite{lloyd2014quantum} proposed a
protocol for quantum principal component analysis (qPCA) based on
density-matrix exponentiation: it uses multiple copies of a state \(\rho\)
to construct \(e^{-i\rho t}\) and then extracts the eigenvalues and
eigenvectors of \(\rho\) via quantum phase estimation. Since it approximates
the largest \(D\) eigenvalues and the corresponding
eigenvectors, the Lanczos method is a natural tool for principal component
analysis when applied to the density matrix of interest. qPCA requires
\(n=O(1/\epsilon^3)\) copies of \(\rho\) and runs in time \(O(n\log d)\),
which is prohibitive when high precision is required. In contrast, the
circuit depth of our parallel quantum Lanczos algorithm is independent of
the precision and is dominated by that of the input oracle. This advantage
stems from a different input model: whereas qPCA accesses the density matrix
as the system state, our protocol accesses it by preparing its vectorized
state.

\section{Discussion}

In this paper, we propose a quantum algorithm for multivariate trace
estimation that features shallow circuit depth at the expense of a large
circuit size. To address the issue of large qubit count, we further provide a
distributed implementation of the algorithmic circuit, yielding a parallel
algorithm for multivariate traces. Our algorithm provides a tradeoff between circuit depth and sampling complexity. Under a fixed coherence-time constraint, computations that were previously
too deep to be executable become feasible. The Frobenius norm factor in the sampling
complexity may limit the applicability of our approach. Nevertheless, our
main contribution is a methodology for designing parallel quantum
algorithms, which is of independent interest and is expected to motivate
algorithm design for other problems. Moreover, since circuit executions can be performed simultaneously, the
sampling complexity does not reflect the actual running time. Suppose there are in total \(M\) circuit executions, each execution takes at most \(\tau_Q\) time, and the statistical aggregation and classical post-processing times are \(T_{\mathrm{aggregate}}\) and \(T_{\mathrm{post}}\), respectively. With \(P\) units capable of independently executing a complete circuit, under ideal scheduling we have
\[
T_Q(P)\lesssim
\left\lceil\frac{M}{P}\right\rceil\tau_Q
+T_{\mathrm{aggregate}}
+T_{\mathrm{post}}.
\]

When applied to spectral sum estimation and quantum Lanczos implementation,
our algorithm does not rely on constructing polynomial transformations of
matrices and provides a reduction in circuit depth by a factor equal to the
degree of the polynomial involved, assuming that the state-preparation input
oracle and the block-encoding oracle have circuit depths of the same order.
We expect our algorithm to find further applications, as multivariate traces
are ubiquitous in quantum information science
\cite{RevModPhys.77.259,schollwck2011densitymatrix,ors2014practical}; we
leave this direction for future work.

Our algorithm is most useful for structured matrices that admit efficient
access through the state-preparation input model, since the circuit depth of
the input oracle dominates the runtime of our algorithm. An avenue for
future work is to identify more classes of matrices whose vectorized states
can be prepared efficiently, beyond those given in this paper.


\bibliographystyle{unsrt}
\bibliography{Ref}

@article{berezutskii2025tensor,
  title={Tensor networks for quantum computing},
  author={Berezutskii, Aleksandr and Liu, Minzhao and Acharya, Atithi and Ellerbrock, Roman and Gray, Johnnie and Haghshenas, Reza and He, Zichang and Khan, Abid and Kuzmin, Viacheslav and Lyakh, Dmitry and others},
  journal={Nature Reviews Physics},
  volume={7},
  number={10},
  pages={581--593},
  year={2025},
  publisher={Nature Publishing Group UK London}
}

@inproceedings{gokhale2021quantum,
  title={Quantum fan-out: Circuit optimizations and technology modeling},
  author={Gokhale, Pranav and Koretsky, Samantha and Huang, Shilin and Majumder, Swarnadeep and Drucker, Andrew and Brown, Kenneth R and Chong, Frederic T},
  booktitle={2021 IEEE International Conference on Quantum Computing and Engineering (QCE)},
  pages={276--290},
  year={2021},
  organization={IEEE}
}

@inproceedings{gilyen2019quantum,
  title={Quantum singular value transformation and beyond: exponential improvements for quantum matrix arithmetics},
  author={Gily{\'e}n, Andr{\'a}s and Su, Yuan and Low, Guang Hao and Wiebe, Nathan},
  booktitle={Proceedings of the 51st annual ACM SIGACT symposium on theory of computing},
  pages={193--204},
  year={2019}
}

@article{mor2025multivariate,
  title={Multivariate trace estimation using quantum state space linear algebra},
  author={Mor-Yosef, Liron and Ubaru, Shashanka and Horesh, Lior and Avron, Haim},
  journal={SIAM Journal on Matrix Analysis and Applications},
  volume={46},
  number={1},
  pages={172--209},
  year={2025},
  publisher={SIAM}
}

@article{zhang2022quantum,
  title={Quantum state preparation with optimal circuit depth: Implementations and applications},
  author={Zhang, Xiao-Ming and Li, Tongyang and Yuan, Xiao},
  journal={Physical Review Letters},
  volume={129},
  number={23},
  pages={230504},
  year={2022},
  publisher={APS}
}

@inproceedings{abramsky2004categorical,
  title={A categorical semantics of quantum protocols},
  author={Abramsky, Samson and Coecke, Bob},
  booktitle={Proceedings of the 19th Annual IEEE Symposium on Logic in Computer Science, 2004.},
  pages={415--425},
  year={2004},
  organization={IEEE}
}

@article{selinger2007dagger,
  title={Dagger compact closed categories and completely positive maps},
  author={Selinger, Peter},
  journal={Electronic Notes in Theoretical computer science},
  volume={170},
  pages={139--163},
  year={2007},
  publisher={Elsevier}
}

@incollection{selinger2010survey,
  title={A survey of graphical languages for monoidal categories},
  author={Selinger, Peter},
  booktitle={New structures for physics},
  pages={289--355},
  year={2010},
  publisher={Springer}
}

@misc{rammal2026orthogonalquantumkrylovdiagonalisation,
      title={Orthogonal Quantum Krylov Diagonalisation}, 
      author={Hadi Rammal and Alexandre Perrin and Oumaya Ladhari and Clément Dutreix and Jérémie Messud and Matthieu Saubanere},
      year={2026},
      eprint={2607.09476},
      archivePrefix={arXiv},
      primaryClass={quant-ph},
      url={https://arxiv.org/abs/2607.09476}, 
}

@article{doi:10.1137/0717059,
author = {Saad, Y.},
title = {On the Rates of Convergence of the Lanczos and the Block-Lanczos Methods},
journal = {SIAM Journal on Numerical Analysis},
volume = {17},
number = {5},
pages = {687-706},
year = {1980},
doi = {10.1137/0717059},
URL = {https://doi.org/10.1137/0717059},
eprint = {https://doi.org/10.1137/0717059}
}

@article{lloyd2014quantum,
  title={Quantum principal component analysis},
  author={Lloyd, Seth and Mohseni, Masoud and Rebentrost, Patrick},
  journal={Nature physics},
  volume={10},
  number={9},
  pages={631--633},
  year={2014},
  publisher={Nature Publishing Group UK London}
}

@article{Kirby_2023,
   title={Exact and efficient Lanczos method on a quantum computer},
   volume={7},
   ISSN={2521-327X},
   url={http://dx.doi.org/10.22331/q-2023-05-23-1018},
   DOI={10.22331/q-2023-05-23-1018},
   journal={Quantum},
   publisher={Verein zur Forderung des Open Access Publizierens in den Quantenwissenschaften},
   author={Kirby, William and Motta, Mario and Mezzacapo, Antonio},
   year={2023},
   month=May, pages={1018} }

@article{yoshioka2025krylov,
  title={Krylov diagonalization of large many-body Hamiltonians on a quantum processor},
  author={Yoshioka, Nobuyuki and Amico, Mirko and Kirby, William and Jurcevic, Petar and Dutt, Arkopal and Fuller, Bryce and Garion, Shelly and Haas, Holger and Hamamura, Ikko and Ivrii, Alexander and others},
  journal={Nature Communications},
  volume={16},
  number={1},
  pages={5014},
  year={2025},
  publisher={Nature Publishing Group UK London}
}

@article{motta2020determining,
  title={Determining eigenstates and thermal states on a quantum computer using quantum imaginary time evolution},
  author={Motta, Mario and Sun, Chong and Tan, Adrian TK and O’Rourke, Matthew J and Ye, Erika and Minnich, Austin J and Brandao, Fernando GSL and Chan, Garnet Kin-Lic},
  journal={Nature Physics},
  volume={16},
  number={2},
  pages={205--210},
  year={2020},
  publisher={Nature Publishing Group UK London}
}

@inproceedings{luongo,
  title = {Quantum Algorithms for Spectral Sums},
  author={Luongo, Alessandro and Shao, Changpeng},
  booktitle={Proceedings of the AAAI Conference on Artificial Intelligence},
  volume={40(29)},
  pages={24178–24188},
  year={2026}
}

@article{PhysRevLett.88.217901,
  title = {Direct Estimations of Linear and Nonlinear Functionals of a Quantum State},
  author = {Ekert, Artur K. and Alves, Carolina Moura and Oi, Daniel K. L. and Horodecki, Micha{\l{}} and Horodecki, Pawe{\l{}} and Kwek, L. C.},
  journal = {Phys. Rev. Lett.},
  volume = {88},
  issue = {21},
  pages = {217901},
  numpages = {4},
  year = {2002},
  month = {May},
  publisher = {American Physical Society},
  doi = {10.1103/PhysRevLett.88.217901},
  url = {https://link.aps.org/doi/10.1103/PhysRevLett.88.217901}
}

@article{2017Entanglement,
  title={Entanglement spectroscopy on a quantum computer},
  author={Johri, Sonika  and  Steiger, Damian S.  and  Troyer, Matthias},
  journal={Phys. Rev. B},
  volume={96},
  number={19},
  pages={195136},
  year={2017},
}

@article{PhysRevLett.120.050406,
  title = {R\'enyi Entropies from Random Quenches in Atomic Hubbard and Spin Models},
  author = {Elben, A. and Vermersch, B. and Dalmonte, M. and Cirac, J. I. and Zoller, P.},
  journal = {Phys. Rev. Lett.},
  volume = {120},
  issue = {5},
  pages = {050406},
  numpages = {6},
  year = {2018},
  month = {Feb},
  publisher = {American Physical Society},
  doi = {10.1103/PhysRevLett.120.050406},
  url = {https://link.aps.org/doi/10.1103/PhysRevLett.120.050406}
}

@article{Pawe2002Method,
  title={Method for Direct Detection of Quantum Entanglement},
  author={Paweł Horodecki and  Ekert, Artur },
  journal={Phys. Rev. Lett},
  volume={89},
  number={12},
  pages={127902},
  year={2002},
}

@article{doi:10.1126/science.aau4963,
author = {Tiff Brydges  and Andreas Elben  and Petar Jurcevic  and Benoît Vermersch  and Christine Maier  and Ben P. Lanyon  and Peter Zoller  and Rainer Blatt  and Christian F. Roos },
title = {Probing Rényi entanglement entropy via randomized measurements},
journal = {Science},
volume = {364},
number = {6437},
pages = {260-263},
year = {2019},
doi = {10.1126/science.aau4963},
URL = {https://www.science.org/doi/abs/10.1126/science.aau4963}
}

@misc{gottesman2001quantumdigitalsignatures,
      title={Quantum Digital Signatures}, 
      author={Daniel Gottesman and Isaac Chuang},
      year={2001},
      eprint={quant-ph/0105032},
      archivePrefix={arXiv},
      primaryClass={quant-ph},
      url={https://arxiv.org/abs/quant-ph/0105032}, 
}

@article{1756006.1953029,
author = {Rasmussen, Carl Edward and Nickisch, Hannes},
title = {Gaussian Processes for Machine Learning (GPML) Toolbox},
year = {2010},
issue_date = {3/1/2010},
publisher = {JMLR.org},
volume = {11},
issn = {1532-4435},
journal = {J. Mach. Learn. Res.},
month = dec,
pages = {3011–3015},
numpages = {5}
}

@InProceedings{pmlr-v119-kerenidis20a,
  title = 	 {Quantum Expectation-Maximization for {G}aussian mixture models},
  author =       {Kerenidis, Iordanis and Luongo, Alessandro and Prakash, Anupam},
  booktitle = 	 {Proceedings of the 37th International Conference on Machine Learning},
  pages = 	 {5187--5197},
  year = 	 {2020},
  editor = 	 {III, Hal Daumé and Singh, Aarti},
  volume = 	 {119},
  series = 	 {Proceedings of Machine Learning Research},
  month = 	 {13--18 Jul},
  publisher =    {PMLR},
  url = 	 {https://proceedings.mlr.press/v119/kerenidis20a.html}
}

@book{rue2005gaussian, 
title={Gaussian Markov Random Fields}, 
ISBN={9780203492024}, 
url={http://dx.doi.org/10.1201/9780203492024}, DOI={10.1201/9780203492024}, 
publisher={Chapman and Hall/CRC}, 
author={Rue, Havard and Held, Leonhard}, 
year={2005}, 
month=Feb }

@article{somma2026quantum, 
title={Quantum algorithm for linear matrix equations}, 
ISSN={2056-6387}, 
url={http://dx.doi.org/10.1038/s41534-026-01360-6}, 
DOI={10.1038/s41534-026-01360-6}, 
journal={npj Quantum Information}, 
publisher={Springer Science and Business Media LLC}, 
author={Somma, Rolando D. and Low, Guang Hao and Berry, Dominic W. and Babbush, Ryan}, 
year={2026}, 
month=Aug 
}

@inproceedings{2999325.2999396,
author = {Hardt, Moritz and Ligett, Katrina and McSherry, Frank},
title = {A simple and practical algorithm for differentially private data release},
year = {2012},
publisher = {Curran Associates Inc.},
address = {Red Hook, NY, USA},
booktitle = {Proceedings of the 26th International Conference on Neural Information Processing Systems - Volume 2},
pages = {2339–2347},
numpages = {9},
location = {Lake Tahoe, Nevada},
series = {NIPS'12}
}

@article{golub1997generalized, 
title={Generalized Cross-Validation for Large-Scale Problems}, 
volume={6}, 
ISSN={1537-2715}, 
url={http://dx.doi.org/10.1080/10618600.1997.10474725}, 
DOI={10.1080/10618600.1997.10474725}, 
number={1}, 
journal={Journal of Computational and Graphical Statistics}, publisher={Informa UK Limited}, 
author={Golub, Gene H. and von Matt, Urs}, 
year={1997}, 
month=Mar, 
pages={1–34} }

@article{dashti2011uncertainty, 
title={Uncertainty Quantification and Weak Approximation of an Elliptic Inverse Problem}, 
volume={49}, ISSN={1095-7170},
url={http://dx.doi.org/10.1137/100814664}, 
DOI={10.1137/100814664}, 
number={6}, 
journal={SIAM Journal on Numerical Analysis}, 
publisher={Society for Industrial & Applied Mathematics (SIAM)}, author={Dashti, M. and Stuart, A. M.}, 
year={2011}, 
month=Jan, 
pages={2524–2542}}

@InProceedings{pmlr-v37-hana15,
  title = 	 {Large-scale log-determinant computation through stochastic Chebyshev expansions},
  author = 	 {Han, Insu and Malioutov, Dmitry and Shin, Jinwoo},
  booktitle = 	 {Proceedings of the 32nd International Conference on Machine Learning},
  pages = 	 {908--917},
  year = 	 {2015},
  editor = 	 {Bach, Francis and Blei, David},
  volume = 	 {37},
  series = 	 {Proceedings of Machine Learning Research},
  address = 	 {Lille, France},
  month = 	 {07--09 Jul},
  publisher =    {PMLR},
  url = 	 {https://proceedings.mlr.press/v37/hana15.html}
}

@inproceedings{3524938.3525419,
author = {Kerenidis, Iordanis and Luongo, Alessandro and Prakash, Anupam},
title = {Quantum expectation-maximization for Gaussian mixture models},
year = {2020},
publisher = {JMLR.org},
booktitle = {Proceedings of the 37th International Conference on Machine Learning},
articleno = {481},
numpages = {11},
series = {ICML'20}
}

@article{RevModPhys.77.259,
  title = {The density-matrix renormalization group},
  author = {Schollw{\"o}ck, U.},
  journal = {Rev. Mod. Phys.},
  volume = {77},
  issue = {1},
  pages = {259--315},
  numpages = {0},
  year = {2005},
  month = {Apr},
  publisher = {American Physical Society},
  doi = {10.1103/RevModPhys.77.259},
  url = {https://link.aps.org/doi/10.1103/RevModPhys.77.259}
}

@article{schollwck2011densitymatrix, 
title={The density-matrix renormalization group in the age of matrix product states}, 
volume={326}, 
ISSN={0003-4916}, 
url={http://dx.doi.org/10.1016/j.aop.2010.09.012}, 
DOI={10.1016/j.aop.2010.09.012}, 
number={1}, 
journal={Annals of Physics}, 
publisher={Elsevier BV}, 
author={Schollwöck, Ulrich}, 
year={2011}, 
month=Jan, 
pages={96–192} }

@article{ors2014practical, 
title={A practical introduction to tensor networks: Matrix product states and projected entangled pair states}, 
volume={349}, ISSN={0003-4916}, 
url={http://dx.doi.org/10.1016/j.aop.2014.06.013}, 
DOI={10.1016/j.aop.2014.06.013}, 
journal={Annals of Physics}, 
publisher={Elsevier BV}, 
author={Orús, Román}, 
year={2014}, 
month=Oct, 
pages={117–158} }

@inproceedings{10.1145/1132516.1132579,
author = {Aharonov, Dorit and Jones, Vaughan and Landau, Zeph},
title = {A polynomial quantum algorithm for approximating the Jones polynomial},
year = {2006},
isbn = {1595931341},
publisher = {Association for Computing Machinery},
address = {New York, NY, USA},
url = {https://doi.org/10.1145/1132516.1132579},
doi = {10.1145/1132516.1132579},
booktitle = {Proceedings of the Thirty-Eighth Annual ACM Symposium on Theory of Computing},
pages = {427–436},
numpages = {10},
location = {Seattle, WA, USA},
series = {STOC '06}
}

@misc{yimsiriwattana2004generalizedghzstatesdistributed,
      title={Generalized GHZ States and Distributed Quantum Computing}, 
      author={Anocha Yimsiriwattana and Samuel J. Lomonaco Jr},
      year={2004},
      eprint={quant-ph/0402148},
      archivePrefix={arXiv},
      primaryClass={quant-ph},
      url={https://arxiv.org/abs/quant-ph/0402148}, 
}

@misc{cuccaro2004newquantumripplecarryaddition,
      title={A new quantum ripple-carry addition circuit}, 
      author={Steven A. Cuccaro and Thomas G. Draper and Samuel A. Kutin and David Petrie Moulton},
      year={2004},
      eprint={quant-ph/0410184},
      archivePrefix={arXiv},
      primaryClass={quant-ph},
      url={https://arxiv.org/abs/quant-ph/0410184}, 
}

@article{ren2020general, title={A general automatic method for optimal construction of matrix product operators using bipartite graph theory}, volume={153}, ISSN={1089-7690}, url={http://dx.doi.org/10.1063/5.0018149}, DOI={10.1063/5.0018149}, number={8}, journal={The Journal of Chemical Physics}, publisher={AIP Publishing}, author={Ren, Jiajun and Li, Weitang and Jiang, Tong and Shuai, Zhigang}, year={2020}, month=Aug}

@article{zhang2024circuit, title={Circuit complexity of quantum access models for encoding classical data}, volume={10}, ISSN={2056-6387}, url={http://dx.doi.org/10.1038/s41534-024-00835-8}, DOI={10.1038/s41534-024-00835-8}, number={1}, journal={npj Quantum Information}, publisher={Springer Science and Business Media LLC}, author={Zhang, Xiao-Ming and Yuan, Xiao}, year={2024}, month=Apr
}

@article{2011832.2011833,
author = {Perez-Garcia, D. and Verstraete, F. and Wolf, M. M. and Cirac, J. I.},
title = {Matrix product state representations},
year = {2007},
issue_date = {July 2007},
publisher = {Rinton Press, Incorporated},
address = {Paramus, NJ},
volume = {7},
number = {5},
issn = {1533-7146},
journal = {Quantum Info. Comput.},
month = jul,
pages = {401–430},
numpages = {30}
}

@article{PhysRevLett.95.110503,
  title = {Sequential Generation of Entangled Multiqubit States},
  author = {Sch{\"o}n, C. and Solano, E. and Verstraete, F. and Cirac, J. I. and Wolf, M. M.},
  journal = {Phys. Rev. Lett.},
  volume = {95},
  issue = {11},
  pages = {110503},
  numpages = {4},
  year = {2005},
  month = {Sep},
  publisher = {American Physical Society},
  doi = {10.1103/PhysRevLett.95.110503},
  url = {https://link.aps.org/doi/10.1103/PhysRevLett.95.110503}
}

@article{shende2006synthesis, 
title={Synthesis of quantum-logic circuits}, 
volume={25}, ISSN={1937-4151}, 
url={http://dx.doi.org/10.1109/TCAD.2005.855930}, 
DOI={10.1109/tcad.2005.855930}, 
number={6}, 
journal={IEEE Transactions on Computer-Aided Design of Integrated Circuits and Systems}, 
publisher={Institute of Electrical and Electronics Engineers (IEEE)}, 
author={Shende, V.V. and Bullock, S.S. and Markov, I.L.}, 
year={2006}, 
month={June}, 
pages={1000–1010}
}

\newpage
\appendix

\section{Matrix Classes with Efficient State-Preparation Input Circuits}\label{app:class_mat}

In this section, we introduce matrix classes whose state-preparation input oracles can be constructed efficiently. Throughout this section, subscripts $C$ and $R$ denote the column register and the row register, respectively. Let $d$ be the circuit depth and $m$ the ancilla-qubit overhead.

\paragraph*{Pauli string.}
\[
P=P_1\otimes\cdots\otimes P_q,\qquad P_k\in\{I,X,Y,Z\}.
\]
Under the convention of column-major vectorization,
\begin{equation}
\vecket{P} = \frac{\operatorname{vec}(P)}{\sqrt{2^q}}
=(I_C\otimes P_R)|\Phi_q\rangle,\qquad
|\Phi_q\rangle=\frac1{\sqrt{2^q}}\sum_j|j\rangle_C|j\rangle_R.
\end{equation}
Therefore, it suffices to use two layers of gates to prepare \(q\) Bell pairs, followed by a layer of single-qubit Pauli gates acting on \(R\). We have
\begin{equation}
d\le3,\qquad m=0.
\end{equation}

\paragraph*{Linear combination of Pauli strings.}
For $A = \sum_{\ell=1}^L \alpha_{\ell} P_{\ell}, L=\mathrm{poly}(q)$ given as a linear combination of Pauli strings, we first present an alternative preparation circuit for Pauli strings. Use two bits $(z,x)$ to encode the single-qubit Pauli operator:
\[
P(z,x)=\mathrm{i}^{zx}X^xZ^z.
\]
Define a two-qubit gate
\begin{equation}
V=\mathrm{CNOT}_{C\to R}(H_C\otimes I_R)\,
\operatorname{diag}(1,1,1,\mathrm{i}),
\end{equation}
whose action is
\begin{equation}
    V\ket{zx} = \mathrm{vec}(P(z,x))/\sqrt{2}.
\end{equation}
For a Pauli string $P$, define a label state
\begin{equation}
    |\lambda(P)\rangle=|z_1\cdots z_q\rangle_C|x_1\cdots x_q\rangle_R.
\end{equation}
Then
\begin{equation}
    U_{P}:=\prod_{k=1}^{q}V_{C_kR_k},\qquad U_{P}|\lambda(P)\rangle=\frac{\operatorname{vec}(P)}{\sqrt{2^q}}.
\end{equation}
Let \(c=\sqrt{\sum_\ell|\alpha_\ell|^2}\). Then
\[
\frac{\operatorname{vec}(A)}{\|A\|_F}
=
U_P
\underbrace{\left(
\sum_{\ell=1}^{L}\frac{\alpha_\ell}{c}
|\lambda(P_\ell)\rangle
\right)}_{\text{\(2q\)-qubit, \(L\)-sparse}}.
\]
According to the optimal quantum sparse state preparation algorithm \cite{zhang2022quantum}, 
\begin{equation}
    d = O(\log(qL)) = O(\log q), \qquad m=O(qL\log L)=O(\mathrm{poly}(q)).
\end{equation}

\paragraph*{Matrices with an efficient singular-vector transformation and preparable singular values.}
Suppose $A = U\Sigma V^\dagger = \sum_k \sigma_k|u_k\rangle\langle v_k|$ and efficient circuits are available for \(U\), for \(V\), and for preparing 
\[
|\sigma\rangle=\frac1{\|\sigma\|_2}\sum_k\sigma_k|k\rangle.
\]
Prepare \(|\sigma\rangle_C|0\rangle_R\), copy the computational-basis labels with parallel CNOTs, and apply \(\bar{V}\) to \(C\) and \(U\) to \(R\):
\[
\vecket{A}
=(\bar{V}\otimes U)
\frac{\sum_k\sigma_k|k\rangle_C|k\rangle_R}{\|\sigma\|_2}.
\]
Thus
\begin{equation}
    d \le d_\sigma+1+\max(d_U,d_{\bar{V}}),
\end{equation}
with polynomial ancilla qubits whenever the component circuits use polynomial workspace.

\paragraph*{Circulant matrices with a preparable generating vector.}
\[
A_{ij}=c_{(i-j)\bmod N},\qquad
|c\rangle=\frac{\sum_k c_k|k\rangle}{\|c\|_2}.
\]
Start with \(|+\rangle_C^{\otimes q}|c\rangle_R\) and perform modular addition:
\[
|j\rangle_C|k\rangle_R\longmapsto
|j\rangle_C|j+k\bmod N\rangle_R.
\]
The resulting state is exactly \(|A\rangle\).
Hence
\begin{equation}
d=d_c+O(q),\qquad m=m_c+O(1)
\end{equation}
by using a linear-depth reversible adder \cite{cuccaro2004newquantumripplecarryaddition}.

\paragraph*{Matrix product operators with polynomial bond dimension.}
Suppose \(A\) is given explicitly as an open-boundary MPO:
\[
A_{\mathbf i,\mathbf j}
=v_L^\dagger W_1^{i_1j_1}\cdots W_q^{i_qj_q}v_R,
\]
where $\mathbf i,\mathbf j\in\{0,1\}^q$, the matrix $W_k^{i_kj_k}$ has dimensions $\chi_{k-1}\times\chi_k$, and
\[
 \chi_0=\chi_q=1,\qquad \max_k\chi_k\le\chi.
\]
This representation contains $O(q\chi^2)$ complex parameters. Consequently, $\chi=\operatorname{poly}(q)$ gives a polynomial-size tensor description even when $A$ is dense and has full matrix rank.

Using column vectorization, the target state is
\begin{equation}
 \vecket{A}:=\frac{\operatorname{vec}(A)}{\|A\|_F}
 =\frac1{\|A\|_F}\sum_{\mathbf i,\mathbf j}
 A_{\mathbf i,\mathbf j}\ket{\mathbf j}_C\ket{\mathbf i}_R.
 \label{eq:target}
\end{equation}
Introduce a four-dimensional local index $s_k=(j_k,i_k)$. In the paired register order
\[
 C_1R_1\,C_2R_2\,\cdots\,C_qR_q,
\]
the same coefficients define the MPS
\begin{equation}
 \ket{\Psi_A}
 =\frac1{\|A\|_F}\sum_{s_1,\ldots,s_q}
 W_1^{i_1j_1}\cdots W_q^{i_qj_q}\ket{s_1}\cdots\ket{s_q}.
 \label{eq:paired-mps}
\end{equation}
Its bond dimension is at most $\chi$. If $P_{\mathrm{ord}}$ denotes the permutation from paired order to standard column--row order, then
\begin{equation}
 \vecket{A}=P_{\mathrm{ord}}\ket{\Psi_A}.
 \label{eq:ordering}
\end{equation}
In the preparation circuit, each local operation can act directly on the labeled pair $C_k,R_k$, so the output can already occupy the standard registers.

The preparation follows three standard steps. First, classical canonicalization converts the tensors into isometries while preserving the represented state \cite{2011832.2011833}. Second, these isometries are extended to unitary operations and applied sequentially: a small auxiliary register carries the virtual bond index between sites and becomes disentangled after the final step \cite{PhysRevLett.95.110503}. Finally, each local unitary is compiled into elementary gates \cite{shende2006synthesis}. Combining sequential generation with generic unitary synthesis gives the conservative bounds
\begin{equation}\label{conser_bound}
    d = O(q\chi^2), \qquad m\le \lceil \log_2(\chi) \rceil,
\end{equation}
with additional polylogarithmic overhead for finite-gate-set approximation.

A particularly useful subclass is
\[
A=\sum_{\ell=1}^{L}\alpha_\ell
B_{\ell,1}\otimes\cdots\otimes B_{\ell,q},
\qquad L=\operatorname{poly}(q),
\]
where the \(B_{\ell,k}\) are arbitrary \(2\times2\) matrices. An exact MPO with bond dimension at most $L$ follows from
\begin{equation}
 \begin{aligned}
 W_1^{ij}
 &=\bigl(\alpha_1(B_{1,1})_{ij},\ldots,
          \alpha_L(B_{L,1})_{ij}\bigr),\\
 W_k^{ij}
 &=\mathrm{diag}\!\bigl((B_{1,k})_{ij},\ldots,(B_{L,k})_{ij}\bigr),
 \quad 2\le k\le q-1,\\
 W_q^{ij}
 &=\bigl((B_{1,q})_{ij},\ldots,(B_{L,q})_{ij}\bigr)^{\mathsf T}.
 \end{aligned}
 \label{eq:sum-products-mpo}
\end{equation}
Systematic methods for constructing and compressing MPOs from sum-of-products expressions are discussed in \cite{ren2020general}. Systematic MPO construction from such expressions is discussed in \cite{ren2020general}. Eq.~\eqref{conser_bound} implies the preparation bounds
\begin{equation}
 D=O(qL^2),\qquad m\le\lceil\log_2L\rceil,
\end{equation}
before finite-gate-set compilation.

\section{Conversions Between the Block-Encoding Input Model and the State Preparation Input Model\label{app:conversion}}

Let $A \in \mathbb{C}^{N \times N}$ and $N=2^q$ for simplicity. We use column-major vectorization and write
\begin{equation}
    |A\rrangle \coloneqq \frac{1}{\|A\|_F} \sum_{i,j} A_{ij}\ket{j}\ket{i}, \qquad \|A\|_F = \sqrt{\mathrm{Tr}(A^\dagger A)}.
\end{equation}
\paragraph{State-preparation input model (S).} A unitary $U_A^S$ acting on two $q$-qubit registers that satisfies $U_A^S\ket{0^{2q}} = |A\rrangle$.
\paragraph{Block-encoding input model (B).} A unitary $U_A^B$ acting on $a$ ancilla qubits and a $q$-qubit system is an exact $(\alpha,a,0)$-block-encoding if
\begin{equation}
    (\bra{0^a} \otimes I)U_A^B(\ket{0^a} \otimes I) = A/\alpha, \qquad \alpha \ge \|A\|_2.
\end{equation}

\subsection{Reduction from block-encoding to state-preparation}
\begin{proposition}
Given $U_A^B$, let registers 1, 2, 3 have dimensions $2^a,2^q,2^q$, respectively, with register 1 serving as the ancillary register for $U_A^B$. Let $\ket{\Phi} = \frac{1}{\sqrt{2^q}} \sum_j \ket{j}\ket{j}$ be prepared on registers 2 and 3. Applying $U_A^B$ to register
3, the second register of $\ket{\Phi}$, with register 1
initialized to \(\ket{0^a}\) produces
\begin{equation}
    \sqrt{p}\ket{0^a}|A\rrangle + \ket{\mathrm{bad}},
\end{equation}
where $(\ket{0^a}\bra{0^a} \otimes I)\ket{\mathrm{bad}}=0$ and $p=\frac{\|A\|_F^2}{\alpha^22^q}$. Postselecting the all-zero state on the ancilla qubits prepares precisely $|A\rrangle$.
\end{proposition}
\begin{proof}
The upper-left block of $U_A^B$, $(\bra{0^a}\otimes I)U_A^B(\ket{0^a}\otimes I)$, acts on register 3 as \(A/\alpha\). Hence, when applied to $\ket{\Phi}$ on registers 2 and 3, it produces
\begin{equation}
(I \otimes A/\alpha)\ket{\Phi} = \frac{\|A\|_F}{\alpha\sqrt{2^q}}|A\rrangle.
\end{equation}
The squared norm of this ancilla-zero component is \(p\).
\end{proof}

Amplitude amplification can be applied to boost the success probability. If $p$ is known, exact amplitude amplification constructs a unitary that prepares $|A\rrangle$ deterministically with asymptotic query complexity $O(1/\sqrt{p})$. If only a positive lower bound $p_{\min} \le p$ is available, fixed-point amplitude amplification yields an approximate preparation with $O(\log(1/\epsilon)/\sqrt{p_{\min}})$ queries and failure probability at most $\epsilon$.

\subsection{Reduction from state-preparation to block-encoding}
\begin{proposition}
    Given $U_A^S$, let registers 1,2,3 each have dimension $2^q$, with register 3 serving as the block-encoding system. Let $U_\Phi$ act on registers 1 and 2 such that $U_\Phi\ket{0^{2q}}=\ket{\Phi} = \frac{1}{\sqrt{2^q}} \sum_j \ket{j}\ket{j}$. Define
    \begin{equation}
        U_A^B = (U_\Phi^\dagger \otimes I) \mathrm{SWAP}_{2,3} (U_A^S \otimes I).
    \end{equation}
    Then $U_A^B$ is an exact $(\sqrt{2^q}\|A\|_F,2q,0)$-block encoding of $A$, using one query to $U_A^S$.
\end{proposition}
\begin{proof}
    For arbitrary $\ket{v}=\sum_k v_k \ket{k}$,
    \begin{equation}
        \begin{aligned}
            &(\bra{\Phi} \otimes I) \mathrm{SWAP_{2,3}} (|A\rrangle \otimes \ket{v}) \\
            =&\frac{1}{\sqrt{2^q}\|A\|_F} \sum_{i,k} A_{ik}v_k \ket{i} = \frac{1}{\sqrt{2^q}\|A\|_F}A\ket{v}.        \end{aligned}
    \end{equation}
    Since $\bra{0^{2q}}U_\Phi^\dagger = \bra{\Phi}$ on registers 1 and 2, projecting the two ancilla registers of $U_A^B$ onto zero extracts the stated matrix block.
\end{proof}
The normalization factor of this block encoding is $\alpha_S = \sqrt{d}\|A\|_F$; its ratio to the smallest possible normalization factor is $\alpha_S/\|A\|_2 = \sqrt{d \cdot \mathrm{sr}(A)} \in [\sqrt{d},d]$, where $\mathrm{sr}(A) = \|A\|_F^2/\|A\|_2^2$ is the so-called stable rank. This suggests that the normalization factor $\alpha_S$ may be unnecessarily large.

In summary, the models admit explicit two-way reductions but may not be efficiently interchangeable.

\section{Tensor Network Representation\label{app_tensor}}

In this section, we introduce the tensor notation for linear algebraic operations. The rigor of tensor calculus is grounded in category theory, which has been applied to model various quantum information processing tasks, such as the formulation of quantum teleportation \cite{abramsky2004categorical} and completely positive maps \cite{selinger2007dagger}. Category theory comes equipped with a powerful graphical language that is equivalent to its algebraic notation. To be precise, a graphical identity is valid if and only if its algebraic counterpart is valid. This property, known as the coherence theorem \cite{selinger2010survey}, justifies the use of diagrammatic proofs in practice.

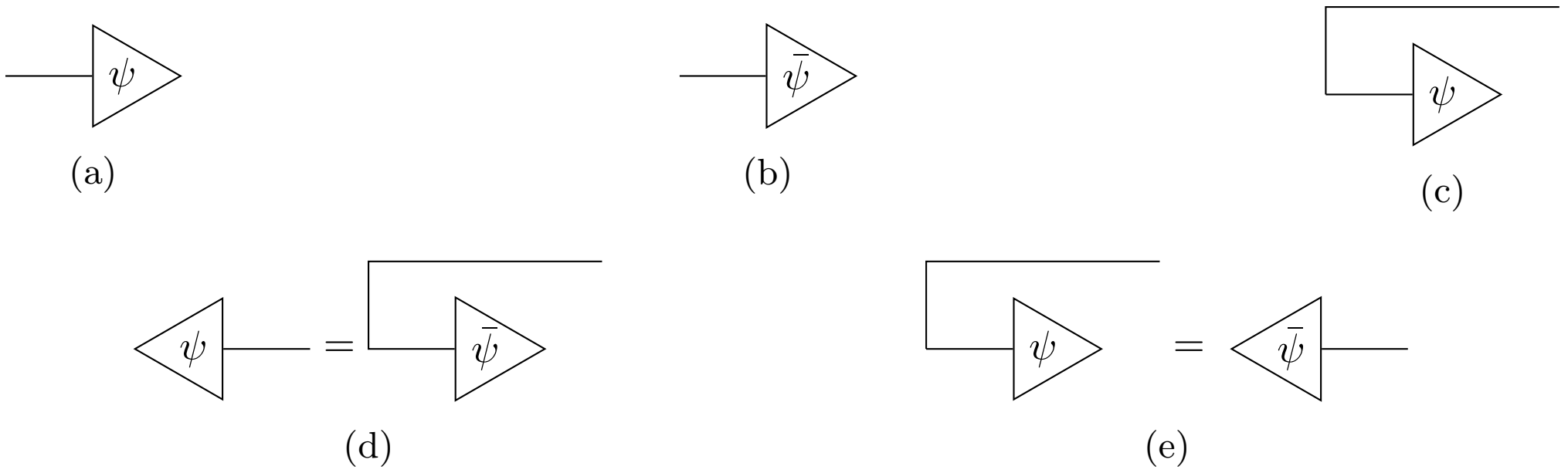
\begin{figure}[htbp]
    \centering
    \begin{minipage}{0.3\textwidth}
    \centering
    \begin{tikzpicture}[
        tensor/.style={draw, triangle right}
    ]
        \node[tensor] (psi) at (0,0) {$\psi$};
        \draw (psi.west) -- (-1,0);
    \end{tikzpicture}
    \subcaption{}
    \end{minipage}
    \hfill
    \begin{minipage}{0.3\textwidth}
    \centering
    \begin{tikzpicture}[
        tensor/.style={draw, triangle right}
    ]
        \node[tensor] (psi) at (0,0) {$\bar{\psi}$};
        \draw (psi.west) -- (-1,0);
    \end{tikzpicture}
    \subcaption{}
    \end{minipage}
    \hfill
    \begin{minipage}{0.3\textwidth}
    \centering
    \begin{tikzpicture}[
        tensor/.style={draw, triangle right}
    ]
        \node[tensor] (psi) at (0,0) {$\psi$};
        \draw (psi.west) -- (-1,0);
        \draw (-1,0) -- (-1,0.75) -- (1,0.75);
    \end{tikzpicture}
    \subcaption{}
    \end{minipage}

    \vspace{1em}

    \centering
    \begin{minipage}{0.3\textwidth}
    \centering
    \begin{tikzpicture}[
        tensor_l/.style={draw, triangle left},
        tensor_r/.style={draw, triangle right}
    ]
        \node[tensor_l] (psi_dagger) at (0,0) {$\psi$};
        \draw (psi_dagger.east) -- (1,0);
    
        \node[draw=none] (op) at (1.25,0) {$=$};
    
        \node[tensor_r] (psi_bar) at (2.5,0) {$\bar{\psi}$};
        \draw (psi_bar.west) -- (1.5,0) -- (1.5,0.75) -- (3.5,0.75);
    \end{tikzpicture}
    \subcaption{}
    \end{minipage}
    \hspace{0.1\textwidth}
    \begin{minipage}{0.3\textwidth}
    \centering
    \begin{tikzpicture}[
        tensor_r/.style={draw, triangle right},
        tensor_l/.style={draw, triangle left}
    ]
        \node[tensor_r] (psi) at (0,0) {$\psi$};
        \draw (psi.west) -- (-1,0);
        \draw (-1,0) -- (-1,0.75) -- (1,0.75);
        
        \node[draw=none] (op) at (1.25,0) {$=$};
    
        \node[tensor_l] (psi_bar) at (2.125,0) {$\bar{\psi}$};
        \draw (psi_bar.east) -- (3.125,0);
    \end{tikzpicture}
    \subcaption{}\label{trans_iden}
    \end{minipage}
    \caption{Tensor diagrams for (a) a ket state $\ket{\psi}$, (b) its complex conjugate, (c) its transpose, (d) its conjugate transpose, and (e) the identity $\ket{\psi}^T=\bra{\bar{\psi}}$.}\label{state_fig}
\end{figure}

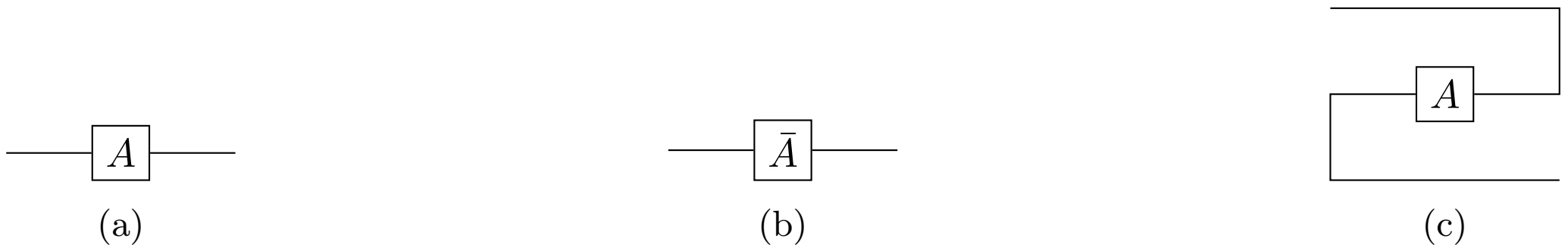
\begin{figure}[htbp]
\centering
\begin{subfigure}{0.3\textwidth}
    \centering
    \begin{tikzpicture}[
        tensor/.style={draw}
    ]
        \node[tensor] (A) at (0,0) {$A$};
        \draw (A.west) -- (-1,0);
        \draw (A.east) -- (1,0);
    \end{tikzpicture}
    \subcaption{}
\end{subfigure}
\hfill
\begin{subfigure}{0.3\textwidth}
    \centering
    \begin{tikzpicture}[
        tensor/.style={draw}
    ]
        \node[tensor] (A) at (0,0) {$\bar{A}$};
        \draw (A.west) -- (-1,0);
        \draw (A.east) -- (1,0);
    \end{tikzpicture}
    \subcaption{}
\end{subfigure}
\hfill
\begin{subfigure}{0.3\textwidth}
    \centering
    \begin{tikzpicture}[
        tensor/.style={draw}
    ]
        \node[tensor] (A) at (0,0) {$A$};
        \draw (A.east) -- (1,0) -- (1,0.75) -- (-1,0.75);
        \draw (A.west) -- (-1,0) -- (-1,-0.75) -- (1,-0.75);
    \end{tikzpicture}
    \subcaption{}
\end{subfigure}
\caption{Tensor diagrams for (a) a matrix $A$, (b) its complex conjugate, and (c) its transpose.}\label{mat_fig}
\end{figure}

We adopt the convention of representing states (vectors) by triangles and linear operators (matrices) by boxes. Figure~\ref{state_fig} displays the tensor diagrams for a ket state $\ket{\psi}$, the complex conjugate of $\ket{\psi}$, the transpose of $\ket{\psi}$, and the conjugate transpose of $\ket{\psi}$ (i.e., the bra state $\bra{\psi}$). Since $\ket{\psi}^T = \bra{\Bar{\psi}}$, the equivalence in Fig.~\ref{trans_iden} follows. The tensor representation of a matrix $A$, together with its complex conjugate and transpose, is illustrated in Fig.~\ref{mat_fig}. 

In the diagram, each open leg corresponds to an index and its orientation determines whether the tensor is a vector, a dual vector, or a linear operator. Specifically, tensors with open legs only on the left represent vectors, those with open legs only on the right represent dual vectors, and tensors with open legs on both sides represent linear operators. For a matrix tensor, the left and right legs correspond to the row and column indices, respectively.

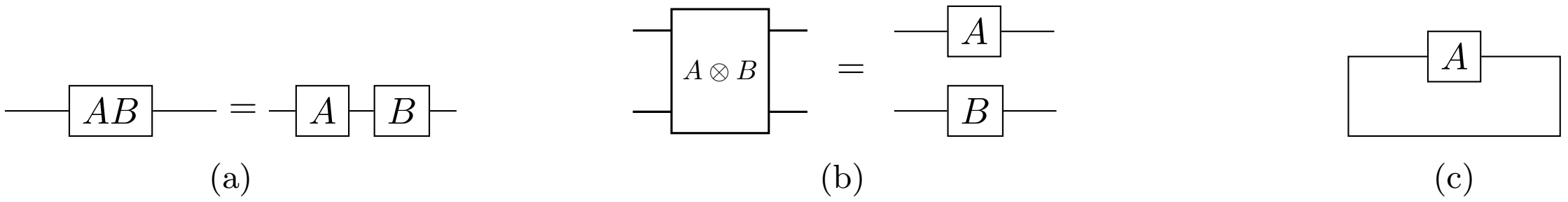
\begin{figure}[htbp]
    \centering
    \begin{subfigure}{0.3\textwidth}
    \centering
    \begin{tikzpicture}[
        tensor/.style={draw}
        ]
            \node[tensor] (A) at (0,0) {$AB$};
            \draw (A.west) -- (-1,0);
            \draw (A.east) -- (1,0);
            \node[draw=none] (op) at (1.25,0) {$=$};
            \node[tensor] (A) at (2,0) {$A$};
            \node[tensor] (B) at (2.75,0) {$B$};
            \draw (A.west)--++(-0.25,0);
            \draw (A.east) -- (B.west);
            \draw (B.east)--++(0.25,0);
        \end{tikzpicture}
    \subcaption{}\label{contra}
    \end{subfigure}
    \hfill
    \begin{subfigure}{0.3\textwidth}
        \centering
        \begin{tikzpicture}[every node/.style={inner sep=0pt}]
            \node[scale=0.73]  (qc) {
            \begin{quantikz}
                 \lstick{} & \gate[2]{A \otimes B} & \qw \\
                 \lstick{} & \qw & \qw
            \end{quantikz}
            };

            \node[draw=none] (op) [right=0.25cm of qc] {$=$};

            \node (tn) [right=0.25cm of op] {
            \begin{tikzpicture}[
            tensor/.style={draw, inner sep=.3333em}
            ]
                \node[tensor] (A) at (0,0) {$A$};
                \node[tensor] (B) at (0,-0.75) {$B$};
                \draw (A.west)--++(-0.5,0);
                \draw (A.east)--++(0.5,0);
                \draw (B.west)--++(-0.5,0);
                \draw (B.east)--++(0.5,0);
            \end{tikzpicture}
            };
        \end{tikzpicture}
        \subcaption{}\label{kron_fig}
    \end{subfigure}
    \hfill
    \begin{subfigure}{0.3\textwidth}
        \centering
        \begin{tikzpicture}[
        tensor/.style={draw}
        ]
            \node[tensor] (A) at (0,0) {$A$};
            \draw (A.west) -- (-1,0) -- (-1,-0.75) -- (1,-0.75) -- (1,0) -- (A.east);
        \end{tikzpicture}
        \subcaption{}\label{tr_fig}
    \end{subfigure}
    \caption{(a) Tensor contraction; (b) Kronecker product; (c) trace.}
\end{figure}

The principal strength of tensor network representations lies in their capacity to compactly express tensors composed of lower-dimensional sub-tensors. Tensor contraction is represented by joining the legs corresponding to the contracted indices, as shown in Fig.~\ref{contra}. The two primary operations that we employ are the Kronecker product and the trace, whose diagrams are given in Figs.~\ref{kron_fig} and~\ref{tr_fig}, respectively.

\begin{figure}[htbp]
\centering
\begin{subfigure}{0.15\textwidth}
    \centering
    \begin{tikzpicture}[
        tensor/.style={draw}
    ]
        \node[tensor] (A) at (0,0) {$A$};
        \draw (A.west) -- (-1,0);
        \draw (A.east) -- (1,0) -- (1,0.75) -- (-1,0.75);
    \end{tikzpicture}
    \subcaption{}\label{vec_fig}
\end{subfigure}
\hspace{0.02\textwidth}
\begin{subfigure}{0.45\textwidth}
    \centering
    \begin{tikzpicture}[
        tensor/.style={draw}
    ]
        \node[tensor] (A) at (0,0) {$A^T$};
        \draw (A.west) -- (-1,0);
        \draw (A.east) -- (1,0) -- (1,0.75) -- (-1,0.75);
    
        \node[draw=none] (op) at (1.25,0.375) {$=$};

        \node[tensor] (A) at (2.5,0.75) {$A$};
        \draw (A.west)--(1.5,0.75);
        \draw (A.east)--(3.5,0.75)--(3.5,0)--(1.5,0);

        \node[draw=none] (op) at (3.75,0.375) {$=$};
    
        \node[tensor] (A) at (5.5,0) {$A$};
        \draw (A.west) -- (5,0) -- (4.5,0.75) -- (4,0.75);
        \draw (A.east) -- (6,0) -- (6,0.75) -- (5,0.75) -- (4.5,0) -- (4,0);
    \end{tikzpicture}
    \subcaption{}\label{vec_trans_fig}
    \label{trans_fig}
\end{subfigure}
\hspace{0.04\textwidth}
\begin{subfigure}{0.3\textwidth}
    \centering
    \begin{tikzpicture}[
        tensor/.style={draw}
    ]
        \node[tensor] (A) at (0,-1.125) {$A$};
        \draw (A.west) -- (-1,-1.125) -- (-1,0.125) -- (1,0.125);
        \draw (A.east) -- (1,-1.125) -- (1,-0.375) -- (-0.75,-0.375) -- (-0.75,0.375) -- (1,0.375);
    
        \node[draw=none] (op) at (1.25,-0.375) {$=$};
    
        \node[tensor] (A) at (2.5,0) {$A$};
        \draw (A.west) -- (1.5,0) -- (1.5,-0.75) -- (3.5,-0.75);
        \draw (A.east) -- (3.5,0);
    \end{tikzpicture}
    \subcaption{}\label{trans_vec_fig}
\end{subfigure}
\caption{Tensor diagrams for (a) $\operatorname{vec}(A)$, (b) $\operatorname{vec}(A^T)$, and (c) $\operatorname{vec}(A)^T$.}
\end{figure}
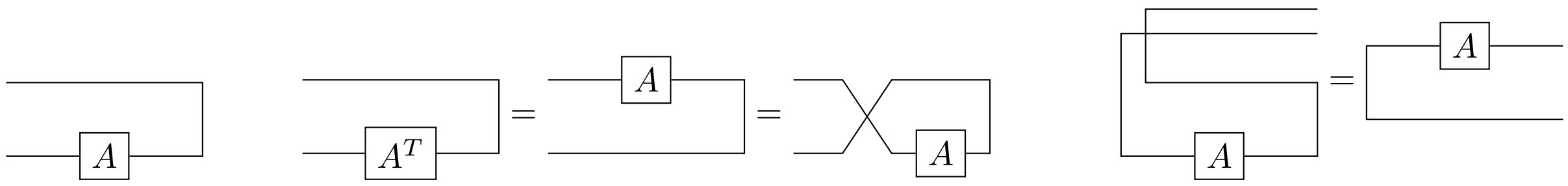

The column-major vectorization of a matrix $A = \sum\limits_{i,j} a_{ij}\ket{i}\bra{j}$ is given by ${\rm vec}(A) = \sum\limits_{i,j}a_{ij}\ket{j}\ket{i}$. In the tensor diagram language, this corresponds to bending the right open leg upward and leftward, as illustrated in Fig.~\ref{vec_fig}. For the transposed matrix, ${\rm vec}(A^T) = \sum\limits_{i,j} a_{ji}\ket{j}\ket{i}=\sum\limits_{i,j}a_{ij}\ket{i}\ket{j}$, which is represented by bending the open right leg of the tensor of $A$ downward and leftward; Fig.~\ref{vec_trans_fig} illustrates how this relates to $\text{vec}(A)$. We also need the transpose of the vectorization ${\rm vec}(A)^T = \sum\limits_{i,j}a_{ij}\bra{j}\bra{i}$, whose diagram is given in Fig.~\ref{trans_vec_fig}.

\begin{figure}
\centering
\begin{subfigure}{0.3\textwidth}
    \centering
    \begin{tikzpicture}[
        tensor/.style={draw}
    ]
        \node[tensor] (A) at (0,0) {$A$};
        \draw (A.west) -- (-1,0);
        \draw (A.east) -- (1,0) -- (1,0.75) -- (-1,0.75);
        \node[tensor] (B) at (0,-0.75) {$B$};
        \draw (B.west) -- (-1,-0.75);
        \draw (B.east) -- (1.25,-0.75) -- (1.25,0.5) -- (-1,0.5);
    
        \node[draw=none] (op) at (1.5,0) {$=$};
    
        \node[tensor] (A) at (3.25,0.25) {$A$};
        \draw (A.west) -- (2.75,0.25) -- (2.25,-0.25) -- (1.75,-0.25);
        \draw (A.east) -- (3.75,0.25) -- (3.75,1) -- (1.75,1);
        \node[tensor] (B) at (3.25,-1) {$B$};
        \draw (B.west) -- (1.75,-1);
        \draw (B.east) -- (3.75,-1) -- (3.75,-0.25) -- (2.75,-0.25) -- (2.25,0.25) -- (1.75,0.25);
    \end{tikzpicture}
    \subcaption{}
\end{subfigure}
\hspace{0.02\textwidth}
\begin{subfigure}{0.3\textwidth}
    \centering
    \begin{tikzpicture}[
        tensor/.style={draw},
        tensor_r/.style={draw, triangle right, minimum size=0.75cm}
    ]
        \node[tensor] (A) at (0,0) {$A$};
        \draw (A.west) -- (-1,0);
        \draw (A.east) -- (1,0) -- (1,0.75) -- (-1,0.75);
        \node[tensor_r] (v) at (0,-0.75) {$v$};
        \draw (v.west) -- (-1,-0.75);
    \end{tikzpicture}
    \subcaption{}
\end{subfigure}
\hspace{0.02\textwidth}
\begin{subfigure}{0.3\textwidth}
    \centering
    \begin{tikzpicture}[
        tensor/.style={draw},
        tensor_r/.style={draw, triangle right, minimum size=0.75cm}
    ]
        \node[tensor] (A) at (0,0) {$A$};
        \draw (A.west) -- (-1,0);
        \draw (A.east) -- (1,0) -- (1,0.75) -- (-1,0.75);
        \node[tensor_r] (v) at (0,-1) {$v$};
        \draw (v.west) -- (-1,-1) -- (-1,-0.5) -- (1.25,-0.5) -- (1.25,0.5) -- (-1,0.5);
    
        \node[draw=none] (op) at (1.5,0) {$=$};
    
        \node[tensor] (A) at (3.25,0) {$A$};
        \draw (A.west) -- (2.75,0) -- (2.25,-0.75) -- (1.75,-0.75);
        \draw (A.east) -- (3.75,0) -- (3.75,0.75) -- (1.75,0.75);
        \node[tensor_r] (v) at (3.25,-0.75) {$v$};
        \draw (v.west) -- (2.75,-0.75) -- (2.25,0) -- (1.75,0);
    \end{tikzpicture}
    \subcaption{}
    \label{vec_row_fig}
\end{subfigure}
\caption{Tensor diagrams for vectorization of Kronecker products: (a) $\operatorname{vec}(A\otimes B)$; (b) $\operatorname{vec}(A\otimes v)$; (c) $\operatorname{vec}(A\otimes v^T)$.}
\label{vec_kron_fig}
\end{figure}
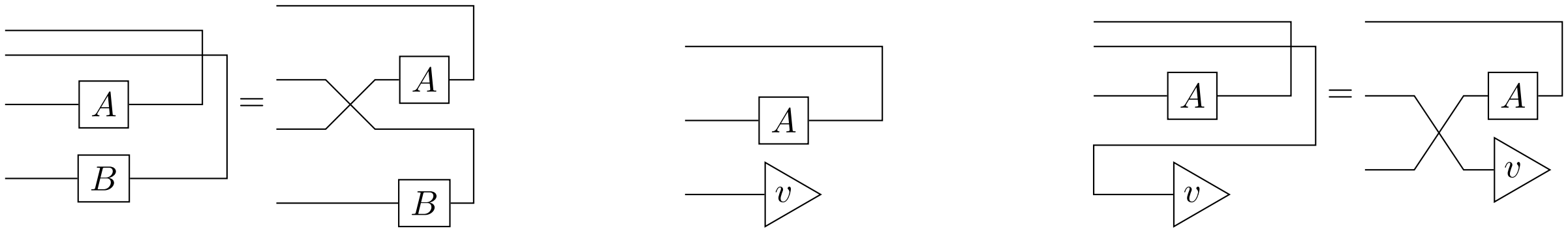

The vectorization of a Kronecker product involving matrices and vectors can be constructed directly from the vectorization of its factors. Let $S_\sigma$ denote the permutation of legs according to the permutation $\sigma$. Figure~\ref{vec_kron_fig} illustrates the following relations.
\begin{align}
    {\rm vec}(A \otimes B) &= S_\sigma {\rm vec}(A)\otimes{\rm vec}(B),\,\sigma=(2,3), \label{vec_kron} \\
    {\rm vec}(A \otimes v) &= {\rm vec}(A) \otimes v, \label{vec_column} \\
    {\rm vec}(A \otimes v^T) &= S_\sigma{\rm vec}(A) \otimes v,\,\sigma=(2,3). \label{vec_row}
\end{align}
Note that $\|A \otimes B\|_F = \|A\|_F\|B\|_F$, hence we have
\begin{align}
    |A \otimes B\rrangle &= S_\sigma |A\rrangle|B\rrangle,\,\sigma = (2,3), \label{prep_kron} \\
    |A \otimes v\rrangle &= |A\rrangle \ket{v}, \label{prep_column} \\
    |A \otimes v^T \rrangle &= S_\sigma |A\rrangle \ket{v},\,\sigma = (2,3). \label{prep_row}
\end{align}

We are now in a position to demonstrate how matrix-vector multiplication can be carried out using the state preparation input model. Assume that we have access to the state preparation oracles for a matrix $A \in \mathbb{C}^{2^p \times 2^q}$ and a vector $v \in \mathbb{C}^{2^q}$, denoted by $U_A$ and $U_v$, respectively. The circuit in Fig.~\ref{mat_vec_qc} encodes $Av$ as a component in the final state:
\begin{equation}
\begin{aligned}
    (U_v^T \otimes I_{2^p}) U_A \ket{0}^{\otimes (p+q)} &= \frac{1}{\|A\|_F} (U_v^T \otimes I_{2^p}) {\rm vec}(A) \\ &= \frac{1}{\|A\|_F}{\rm vec}(AU_v) \\ &= \frac{1}{\|A\|_F}{\rm vec}\left(\left[\frac{Av}{\|v\|_2} \, * \, \cdots \, *\right]\right) \\ &= \ket{0}^{\otimes q} \otimes \frac{Av}{\|A\|_F\|v\|_2} + \ket{\perp},
\end{aligned}
\end{equation}
where $(\bra{0}^{\otimes q} \otimes I_{2^p})\ket{\perp} = 0$ and the second equality follows from Fig.~\ref{mat_vec_tn}. Here and throughout, $I_n$ denotes the $n \times n$ identity matrix.

\begin{figure}
\centering
\begin{subfigure}{0.45\textwidth}
    \centering
    \scalebox{0.75}{
    \begin{quantikz}
        \lstick{$\ket{0}$} & \gate[2]{U_A} & \gate{U_v^T} & \qw \\
        \lstick{$\ket{0}$} & \qw & \qw & \qw 
    \end{quantikz}
    }
    \subcaption{The circuit for matrix-vector multiplication.}
    \label{mat_vec_qc}
\end{subfigure}
\hspace{0.05\textwidth}
\begin{subfigure}{0.45\textwidth}
    \centering
    \begin{tikzpicture}[
    tensor/.style={draw}
    ]
        \node[tensor] (A) at (1.5,0) {$A$};
        \node[tensor] (B) at (0.5,0.75) {$B^T$};
        \draw (A.west) -- (-0.25,0);
        \draw (A.east) -- (2,0) -- (2,0.75) -- (B.east);
        \draw (B.west) -- (-0.25,0.75);

        \node[draw=none] (op) at (2.25,0.375) {$=$};

        \node[tensor] (A) at (3.15,0) {$A$};
        \node[tensor] (B) at (4,0) {$B$};
        \draw (A.west) -- (2.5,0);
        \draw (A.east) -- (B.west);
        \draw (B.east) -- (4.5, 0) -- (4.5, 0.75) -- (2.5, 0.75);
    \end{tikzpicture}
    \subcaption{Tensor diagram for the identity $(B^T\otimes I)\operatorname{vec}(A) = \operatorname{vec}(AB)$.}
    \label{mat_vec_tn}
\end{subfigure}
\caption{}
\end{figure}

\section{Trace Evaluation\label{app_tr_eva}}

In this section, we provide proofs for the lemmas given in the main text.

For odd $k>1$, we define the matrices
\begin{equation}
    \begin{aligned}
        E_{i,k} &= A_{2i}^T \otimes A_{2(k+1-i)},\,i = 1,\dots,\frac{k-1}{2}, \\
        E_{\frac{k+1}{2},k} &= A_{k+1}^T, \\
        O_{i,k} &= A_{2i+1}^T \otimes A_{2(k-i)+1},\,i=1,\dots,\frac{k-1}{2},
    \end{aligned}
\end{equation}
and for even $k>1$, we define the matrices
\begin{equation}
    \begin{aligned}
        E_{i,k} &= A_{2i}^T \otimes A_{2(k+1-i)},\,i = 1,\dots,\frac{k}{2}, \\
        O_{i,k} &= A_{2i+1}^T \otimes A_{2(k-i)+1},\,i=1,\dots,\frac{k}{2}-1, \\
        O_{\frac{k}{2},k} &= A_{k+1}^T.
    \end{aligned}
\end{equation}

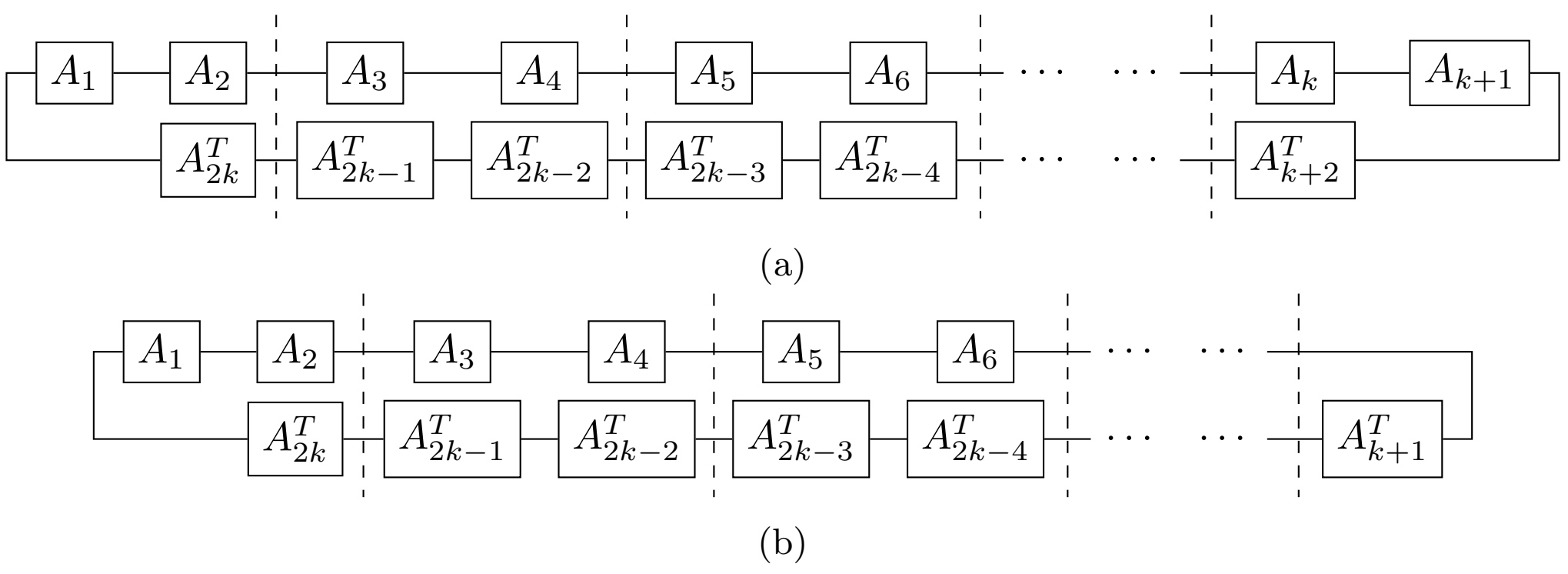
\begin{figure}[htbp]
    \centering
    \begin{subfigure}{\textwidth}
    \centering
    \begin{tikzpicture}[every node/.style={draw}]
        \node (A1) at (0,0) {$A_1$};
        \node (A2) at (1.15,0) {$A_2$};
        \node (A2k) at (1.15,-0.75) {$A_{2k}^T$};
        \node (A3) at (2.5,0) {$A_3$};
        \node (A2k-1) at (2.5,-0.75) {$A_{2k-1}^T$};
        \node (A4) at (4,0) {$A_4$};
        \node (A2k-2) at (4,-0.75) {$A_{2k-2}^T$};
        \node (A5) at (5.5,0) {$A_5$};
        \node (A2k-3) at (5.5,-0.75) {$A_{2k-3}^T$};
        \node (A6) at (7,0) {$A_6$};
        \node (A2k-4) at (7,-0.75) {$A_{2k-4}^T$};
        \node[draw=none] (sym1) at (8.35,0) {$\cdots$};
        \node[draw=none] (sym2) at (8.35,-0.75) {$\cdots$};
        \node[draw=none] (sym3) at (9.15,0) {$\cdots$};
        \node[draw=none] (sym4) at (9.15,-0.75) {$\cdots$};
        \node (Ak) at (10.5,0) {$A_k$};
        \node (Ak+2) at (10.5,-0.75) {$A_{k+2}^T$};
        \node (Ak+1) at (12,0) {$A_{k+1}$};

        \draw (A1.west)--++(-0.25,0)--++(0,-0.75)--(A2k.west);
        \draw (A2k.east)--(A2k-1.west);
        \draw (A2k-1.east)--(A2k-2.west);
        \draw (A2k-2.east)--(A2k-3.west);
        \draw (A2k-3.east)--(A2k-4.west);
        \draw (A2k-4.east)--(sym2.west);
        \draw (sym4.east)--(Ak+2.west);
        \draw (Ak+2.east)--($(Ak+1.east)+(0.25,0)+(0,-0.75)$)--++(0,0.75)--(Ak+1.east);
        \draw (Ak+1.west)--(Ak.east);
        \draw (Ak.west)--(sym3.east);
        \draw (sym1.west)--(A6.east);
        \draw (A6.west)--(A5.east);
        \draw (A5.west)--(A4.east);
        \draw (A4.west)--(A3.east);
        \draw (A3.west)--(A2.east);
        \draw (A2.west)--(A1.east);
        \draw[dashed] ($(A2k.east)!0.5!(A2k-1.west)+(0,1.25)$)--($(A2k.east)!0.5!(A2k-1.west)+(0,-0.5)$);
        \draw[dashed] ($(A2k-2.east)!0.5!(A2k-3.west)+(0,1.25)$)--($(A2k-2.east)!0.5!(A2k-3.west)+(0,-0.5)$);
        \draw[dashed] ($(A2k-4.east)+(0.2,0)+(0,1.25)$)--($(A2k-4.east)+(0.2,0)+(0,-0.5)$);
        \draw[dashed] ($(Ak+2.west)+(-0.2,0)+(0,1.25)$)--($(Ak+2.west)+(-0.2,0)+(0,-0.5)$);
    \end{tikzpicture}
    \subcaption{}
    \end{subfigure}
    \begin{subfigure}{\textwidth}
    \centering
    \begin{tikzpicture}[every node/.style={draw}]
        \node (A1) at (0,0) {$A_1$};
        \node (A2) at (1.15,0) {$A_2$};
        \node (A2k) at (1.15,-0.75) {$A_{2k}^T$};
        \node (A3) at (2.5,0) {$A_3$};
        \node (A2k-1) at (2.5,-0.75) {$A_{2k-1}^T$};
        \node (A4) at (4,0) {$A_4$};
        \node (A2k-2) at (4,-0.75) {$A_{2k-2}^T$};
        \node (A5) at (5.5,0) {$A_5$};
        \node (A2k-3) at (5.5,-0.75) {$A_{2k-3}^T$};
        \node (A6) at (7,0) {$A_6$};
        \node (A2k-4) at (7,-0.75) {$A_{2k-4}^T$};
        \node[draw=none] (sym1) at (8.35,0) {$\cdots$};
        \node[draw=none] (sym2) at (8.35,-0.75) {$\cdots$};
        \node[draw=none] (sym3) at (9.15,0) {$\cdots$};
        \node[draw=none] (sym4) at (9.15,-0.75) {$\cdots$};
        \node (Ak+1) at (10.5,-0.75) {$A_{k+1}^T$};

        \draw (A1.west)--++(-0.25,0)--++(0,-0.75)--(A2k.west);
        \draw (A2k.east)--(A2k-1.west);
        \draw (A2k-1.east)--(A2k-2.west);
        \draw (A2k-2.east)--(A2k-3.west);
        \draw (A2k-3.east)--(A2k-4.west);
        \draw (A2k-4.east)--(sym2.west);
        \draw (sym4.east)--(Ak+1.west);
        \draw (Ak+1.east)--++(0.25,0)--++(0,0.75)--(sym3.east);
        \draw (sym1.west)--(A6.east);
        \draw (A6.west)--(A5.east);
        \draw (A5.west)--(A4.east);
        \draw (A4.west)--(A3.east);
        \draw (A3.west)--(A2.east);
        \draw (A2.west)--(A1.east);
        \draw[dashed] ($(A2k.east)!0.5!(A2k-1.west)+(0,1.25)$)--($(A2k.east)!0.5!(A2k-1.west)+(0,-0.5)$);
        \draw[dashed] ($(A2k-2.east)!0.5!(A2k-3.west)+(0,1.25)$)--($(A2k-2.east)!0.5!(A2k-3.west)+(0,-0.5)$);
        \draw[dashed] ($(A2k-4.east)+(0.2,0)+(0,1.25)$)--($(A2k-4.east)+(0.2,0)+(0,-0.5)$);
        \draw[dashed] ($(Ak+1.west)+(-0.2,0)+(0,1.25)$)--($(Ak+1.west)+(-0.2,0)+(0,-0.5)$);
    \end{tikzpicture}
    \subcaption{}
    \end{subfigure}
    \caption{Tensor network partition for (a) odd $k$ and (b) even $k$, respectively.}
    \label{odd_even}
\end{figure}

\begin{figure}
    \centering
    \begin{subfigure}{0.4\textwidth}
        \centering
        \begin{tikzpicture}[every node/.style={draw}]
        \node (A1) at (-1.1875,0) {$A_1$};
        \node (A2) at (0,0) {$A_2$};
        \node (A6) at (0,-1) {$A_6^T$};
        \node (A3) at (1,-2) {$A_3^T$};
        \node (A4) at (0,-2) {$A_4^T$};
        \node (A5) at (1,-3) {$A_5$};
    
        \draw (A1) -- (A2) -- (1.5,0) -- (1.5,-2) -- (A3);
        \draw (A3) -- (A4) -- (-0.5,-2) -- (-0.5,-3) -- (A5);
        \draw (A5) -- (1.65,-3) -- (1.65,-1) -- (A6) -- (-1.75,-1) -- (-1.75,0) -- (A1);

        \draw[dashed] (-0.625,0.5) -- (-0.625,-3.5);
        \draw[dashed] (0.5,0.5) -- (0.5,-3.5);
        \end{tikzpicture}
        \subcaption{}\label{eg_3}
    \end{subfigure}
    \begin{subfigure}{0.4\textwidth}
        \centering
        \begin{tikzpicture}[every node/.style={draw}]
            \node (A1) at (-1.5625,0) {$A_1$};
            \node (A2) at (0,0) {$A_2$};
            \node (A8) at (0,-1) {$A_8^T$};
            \node (A4) at (0,-2) {$A_4^T$};
            \node (A6) at (0,-3) {$A_6$};
            \node (A3) at (1,-2) {$A_3^T$};
            \node (A7) at (1,-3) {$A_7$};
            \node (A5) at (1,-5) {$A_5^T$};

            \draw (A1) -- (A2) -- (1.5,0) -- (1.5,-2) -- (A3) -- (A4);
            \draw (A4) -- (-0.5,-2) -- (-0.5,-4) -- (1.5,-4) -- (1.5,-5) -- (A5);
            \draw (A5) -- (-0.65,-5) --(-0.65,-3) -- (A6) -- (A7) -- (1.65,-3) -- (1.65,-1) -- (A8) -- (-2.25,-1) -- (-2.25,0) -- (A1);

            \draw[dashed] (0.5,0.5) -- (0.5,-5.5);
            \draw[dashed] (-0.875,0.5) -- (-0.875,-5.5);
        \end{tikzpicture}
        \subcaption{}
    \end{subfigure}
    \caption{Examples for (a) $k=3$ and (b) $k=4$.}\label{eg_fig}
\end{figure}
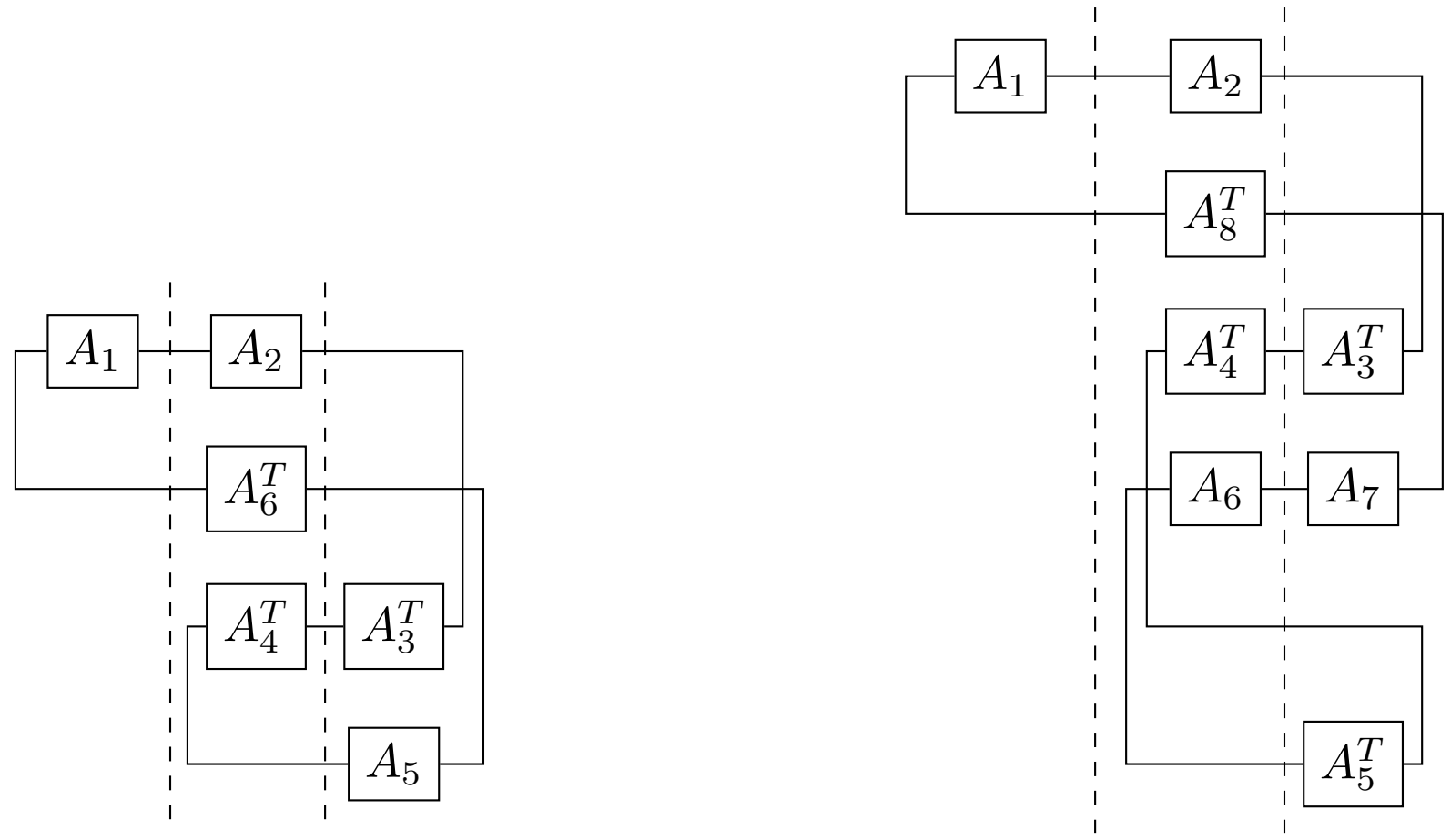

\begin{lemma}[Lemma \ref{tr_Mv} in the main text]
The multivariate trace ${\rm Tr}(A_1A_2 \cdots A_{2k})$ can be expressed as the matrix-vector multiplication $u^TMv$, where $u = {\rm vec}(A_1)$,
\begin{align}
    M &= E_{1,k}^T \otimes {\rm vec}(E_{2,k})^T \otimes \cdots \otimes {\rm vec}(E_{\lceil\frac{k}{2}\rceil,k})^T \label{M_app} \\ 
    v &= {\rm vec}(O_{1,k}) \otimes \cdots \otimes {\rm vec}(O_{\lfloor\frac{k}{2}\rfloor,k}) \label{v_app}
\end{align}
\end{lemma}
\begin{proof}
    We partition the tensor network for $\operatorname{Tr}(A_1A_2\cdots A_{2k})$ into segments as shown in Fig.~\ref{odd_even}, and then transform each segment according to Fig.~\ref{block}. Concatenating the transformed segments in order yields the constructions in Eqs~\eqref{M_app} and \eqref{v_app}. Examples for $k=3$ and $k=4$ are illustrated in Fig.~\ref{eg_fig}.
\end{proof}

\begin{figure}[htbp]
    \centering
    \begin{subfigure}{0.45\textwidth}
        \centering
        \begin{tikzpicture}[every node/.style={draw}]
            \node (A) at (0,0) {$A$};
            \node (B) at (1,0) {$B$};
            \node (C) at (0,-0.75) {$C$};
            \node (D) at (1,-0.75) {$D$};
            \draw (A.west)--++(-0.25,0);
            \draw (C.west)--++(-0.25,0);
            \draw (A.east)--(B.west);
            \draw (B.east)--++(0.25,0);
            \draw (C.east)--(D.west);
            \draw (D.east)--++(0.25,0);
            
            \node[draw=none] (sym) at (2,-0.375) {$=$};
            \node (B) at (3,0) {$B^T$};
            \node (A) at (4,0) {$A^T$};
            \node (D) at (3,-0.75) {$D^T$};
            \node (C) at (4,-0.75) {$C^T$};
            \draw (B.west)--++(-0.15,0)--++(0,-1.5)--($(D)+(0.5,0)+(0,-0.75)$);
            \draw (D.west)--++(-0.25,0)--++(0,-1)--($(D)+(0.5,0)+(0,-1)$);
            \draw (B.east)--(A.west);
            \draw (D.east)--(C.west);
            \draw (A.east)--++(0.15,0)--++(0,1)--($(A)+(-0.5,0)+(0,1)$);
            \draw (C.east)--++(0.25,0)--++(0,1.5)--($(A)+(-0.5,0)+(0,0.75)$);
        \end{tikzpicture}
        \subcaption{}
    \end{subfigure}
    \begin{subfigure}{0.45\textwidth}
        \centering
        \begin{tikzpicture}[every node/.style={draw}]
            \node (A) at (0,0) {$A$};
            \node (B) at (1,0) {$B$};
            \node (C) at (0,-0.75) {$C$};
            \draw (A.west)--++(-0.25,0);
            \draw (C.west)--++(-0.25,0);
            \draw (A.east)--(B.west);
            \draw (B.east)--++(0.25,0)--++(0,-0.75)--(C.east);

            \node[draw=none] (op) at (2,-0.375) {$=$};
            \node (B) at (3,-0.75) {$B^T$};
            \node (A) at (4,-0.75) {$A^T$};
            \node (C) at (3,-1.5) {$C^T$};
            \draw (B.west)--++(-0.25,0)--++(0,-0.75)--(C.west);
            \draw (C.east)--($(A.east)+(0.25,0)+(0,-0.75)$)--++(0,1.5)--(3.5,0);
            \draw (B.east)--(A.west);
            \draw (A.east)--++(0.15,0)--++(0,1)--(3.5,0.25);
        \end{tikzpicture}
        \subcaption{}
    \end{subfigure}
    \caption{Tensor network transformations for two kinds of segments.}
    \label{block}
\end{figure}
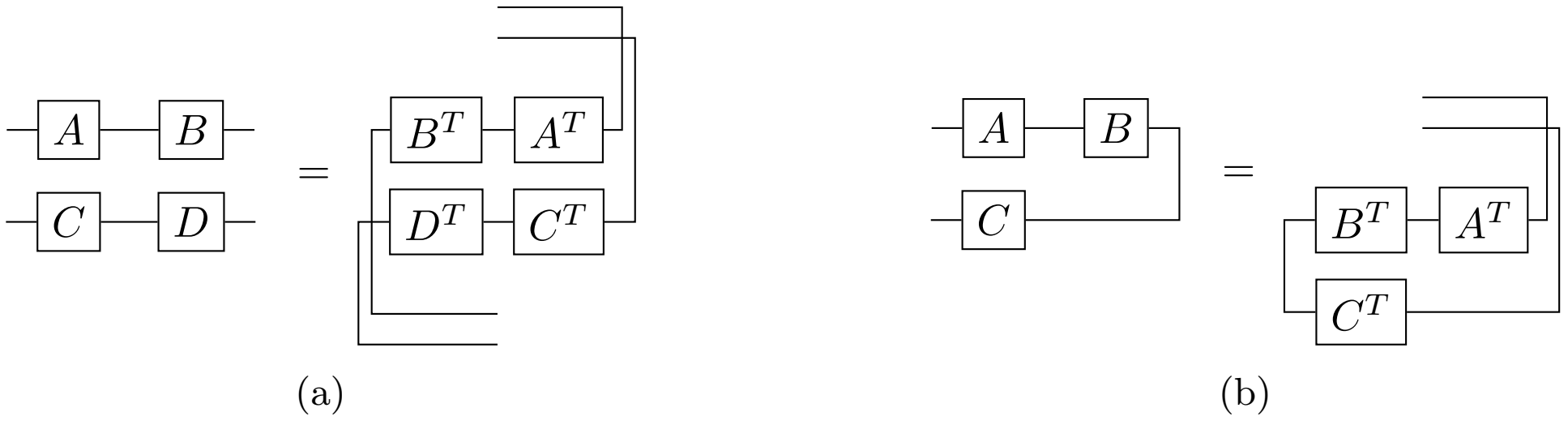

We introduce permutations $\sigma_i = (4i-2,4i-1)$, $\sigma=(2k-1,2k-3,\dots,5,3)(2k,2k-2,\dots,6,4)$, $\tau_{i} = (4i-3,4i-2)$, $\tau=(3,4)$, and $\tau'=(2k-1,2k)$, and denote by $S_i$, $S_\sigma$, $T_i$, $T_\tau$, $T_{\tau'}$ the corresponding permutations of the registers, respectively.

\begin{lemma}[Lemma \ref{input_M_v} in the main text]
    Given the state preparation oracles $U_2,U_3,\dots,U_{2k}$ for matrices $A_2,A_3,\dots,A_{2k}$, define $V_1=U_2 \otimes (T_\tau U_{2k})$, $V_i = (T_iU_{2i}) \otimes U_{2(k+1-i)}$ for $i>1$, and $W_i = (T_i U_{2i+1}) \otimes U_{2(k-i)+1}$. The state preparation unitaries for $M$ and $v$ as defined in Eqs.~\eqref{M_app} and~\eqref{v_app} can be constructed as follows:
    \begin{align}
        U_M &= S_{\sigma} \cdot (S_1V_1) \otimes (S_2 V_2) \otimes \cdots \otimes (S_{\frac{k-1}{2}}V_{\frac{k-1}{2}}) \otimes (T_{\tau'} U_{k+1}), \\
        U_v &= (S_1W_1) \otimes (S_2 W_2) \otimes \cdots \otimes (S_{\frac{k-1}{2}}W_{\frac{k-1}{2}})
    \end{align}
    for odd $k>1$, and
    \begin{align}
        U_M &= S_{\sigma} \cdot (S_1V_1) \otimes (S_2 V_2) \otimes \cdots \otimes (S_{\frac{k}{2}}V_{\frac{k}{2}}) , \\
        U_v &= (S_1W_1) \otimes (S_2 W_2) \otimes \cdots \otimes (S_{\frac{k}{2}-1}W_{\frac{k}{2}-1}) \otimes (T_{\tau'} U_{k+1})
    \end{align}
    for even $k>1$.
\end{lemma}
\begin{proof}
We prove the lemma for odd $k>1$ and for the matrix $M$; the proof for $v$ and for even $k>1$ is similar. Let $w={\rm vec}(E_{2,k}) \otimes \cdots \otimes {\rm vec}(E_{\frac{k+1}{2},k})$. For $M=E_{1,k}^T \otimes w^T$, Eq.~\eqref{vec_row} gives
\begin{equation}
    \operatorname{vec}(M) = S_\sigma \cdot \operatorname{vec}(E_{1,k}^T) \otimes w = S_\sigma \cdot \operatorname{vec}(E_{1,k}^T) \otimes \operatorname{vec}(E_{2,k}) \otimes \cdots \otimes \operatorname{vec}(E_{\frac{k+1}{2},k}),
\end{equation}
where $\sigma = (2k-1,2k-3,\dots,5,3)(2k,2k-2,\dots,6,4)$. Here $S_\sigma$ swaps the row-index register of $E_{1,k}^T$ with that of $w$, as depicted in Fig.~\ref{vec_row_fig}. By Eq.~\eqref{vec_kron},
\begin{align}
    \operatorname{vec}(E_{1,k}^T) &= S_1 \left(\operatorname{vec}(A_2) \otimes \operatorname{vec}(A_{2k}^T) \right),\\
    {\rm vec}(E_{i,k}) &= S_i \left({\rm vec(A_{2i}^T)} \otimes {\rm vec}(A_{2(k+1-i)})\right),\,i = 2,\dots,\frac{k-1}{2},\\
    \operatorname{vec}(E_{\frac{k+1}{2},k}) &= \operatorname{vec}(A_{k+1}^T).
\end{align}
Recalling that vectorizing a transposed matrix amounts to swapping the legs of the row and column indices of $\operatorname{vec}(A)$ (Fig.~\ref{trans_fig}), we obtain 
\begin{align}
    \operatorname{vec}(E_{1,k}^T) &= S_1 \big(\operatorname{vec}(A_2)\otimes (T_\tau\operatorname{vec}(A_{2k}))\big), \\
    {\rm vec}(E_{i,k}) &= S_i \big((T_i{\rm vec(A_{2i})}) \otimes {\rm vec}(A_{2(k+1-i)})\big),\,i = 2,\dots,\frac{k-1}{2}, \\
    {\rm vec}(E_{\frac{k+1}{2},k}) &= T_{\tau'}{\rm vec}(A_{k+1}).
\end{align}
Consequently, 
\begin{equation}
    |M\rrangle = S_\sigma \cdot \big(S_1|A_2\rrangle(T_\tau|A_{2k}\rrangle)\big) \big( S_2(T_2|A_4\rrangle)|A_{2(k-1)}\rrangle \big) \cdots \big( S_\frac{k-1}{2}(T_{\frac{k-1}{2}}|A_{k-1}\rrangle)|A_{k+3}\rrangle \big)\big( T_{\tau'}|A_{k+1}\rrangle \big),
\end{equation}
which yields the circuit construction for $U_M$.
\end{proof}

\begin{lemma}[Lemma \ref{mat_prod_vec} in the main text]
    For matrices $A_2,A_3,\dots,A_{2k} \in \mathbb{C}^{2^q\times 2^q}$, $U_\psi = (U_v^T \otimes I_{2^{2q}}) U_M$ is a $2kq$-qubit quantum circuit that acts on $\ket{0}^{\otimes 2kq}$ as
    \begin{equation}
        \begin{aligned}
            U_\psi \ket{0}^{\otimes 2kq} = 
            & (U_v^T \otimes I_{2^{2q}}) U_M \ket{0}^{\otimes 2kq} \\ =
            & \ket{0}^{\otimes 2(k-1)q} \otimes \frac{\operatorname{vec}(A_{2k}^TA_{2k-1}^T\cdots A_2^T)}{\|A_2\|_F\|A_3\|_F \cdots \|A_{2k}\|_F} + \ket{\perp},
        \end{aligned}
    \end{equation}
    where $(\bra{0}^{\otimes 2(k-1)q}\otimes I_{2^{2q}})\ket{\perp} = 0$.
\end{lemma}
\begin{proof}
    By Lemma \ref{tr_Mv}, $\operatorname{vec}(A_{2k}^TA_{2k-1}^T \cdots A_2^T)$ can be transformed into the product of $M$ and $v$. Following the matrix-vector multiplication procedure developed in the previous section, we construct the circuit 
    \begin{equation}
        U_\psi = \left(U_v^T \otimes I_{2^{2q}}\right) U_M,
    \end{equation}
    where $U_M$ and $U_v$ are the state preparation unitaries for $M$ and $v$, respectively. This circuit acts on $\ket{0}^{\otimes 2kq}$ as
    \begin{equation}
        \begin{aligned}
            U_\psi\ket{0}^{\otimes 2kq} &= \ket{0}^{\otimes 2(k-1)q} \otimes \frac{Mv}{\|A_2\|_F\|A_3\|_F \cdots \|A_{2k}\|_F} + \ket{\perp} \\
            &= \ket{0}^{\otimes 2(k-1)q} \otimes \frac{\operatorname{vec}(A_{2k}^TA_{2k-1}^T \cdots A_2^T)}{\|A_2\|_F\|A_3\|_F \cdots \|A_{2k}\|_F} + \ket{\perp},
        \end{aligned}
    \end{equation}
    where $(\bra{0}^{\otimes 2(k-1)q}\otimes I_{2^{2q}})\ket{\perp} = 0$. 

    Provided that the oracles $U_i$ are implemented using the universal gate set $\{H,S,T,\mathrm{CNOT}\}$, all of whose gates are symmetric under transposition, then $U_i^T$ can be realized by simply reversing the order of the gates.
\end{proof}

\section{The Treatment of the Odd Case and of Observables}\label{app:odd}

\subsection{Treatment of the odd case}\label{sec:odd}
In Section~\ref{multi_trace_est}, we present an algorithm only for a
multivariate trace of an even number of matrix products. For the odd case,
simply appending an identity matrix would reduce the problem to the even
case, but this would introduce a factor of \(2^{q}\) in the complexity,
rendering this naive approach inefficient.

In the quantum implementation of the Lanczos method in
Section~\ref{para_OQKD}, the relevant multivariate traces are of the form
\begin{equation}\label{tr_phi}
    \operatorname{Tr}\!\left(
    A_1A_2\cdots A_k\,\ket{\phi}\bra{\phi}
    \right).
\end{equation}
Assuming access to a state-preparation unitary \(U\) for
\(\ket{\phi}\), i.e., \(U\ket{0}=\ket{\phi}\), the state-preparation
input oracle for the density matrix \(\ket{\phi}\bra{\phi}\) is
\(\bar{U}\otimes U\).
For odd \(k\), Eq.~\eqref{tr_phi} is a multivariate trace of an even number
of matrix products, which is covered by the algorithm in the main text.

\begin{figure}
    \centering
    \includegraphics[width=0.25\linewidth]{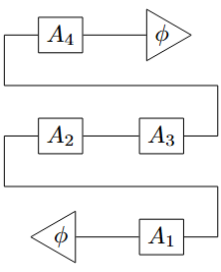}
    \caption{Tensor network representation of $\bra{\phi}A_1A_2A_3A_4\ket{\phi}$ as the inner product $\langle L|R \rangle$.}
    \label{fig:odd}
\end{figure}

For an even number \(2k\) of matrices,
\begin{equation}
    \operatorname{Tr}\!\left(
    A_1A_2\cdots A_{2k}\,\ket{\phi}\bra{\phi}
    \right)
    =
    \bra{\phi}A_1A_2\cdots A_{2k}\ket{\phi}.
\end{equation}
We define the states
\begin{equation}\label{LR}
    \begin{aligned}
        & \ket{R} = \ket{\phi} \otimes \vecket{A_{2k-1}} \otimes \vecket{A_{2k-3}} \otimes \cdots \otimes \vecket{A_1}, \\
        & \ket{L} = \vecket{\bar{A}_{2k}} \otimes \vecket{\bar{A}_{2k-2}} \otimes \cdots \otimes \vecket{\bar{A}_2} \otimes \ket{\phi}.
    \end{aligned}
\end{equation}
Then 
\begin{equation}
    \bra{L}R\rangle = \frac{\bra{\phi} A_1A_2 \cdots A_{2k}\ket{\phi}}{\prod_{j=1}^{2k}\|A_j\|_F}.
\end{equation}
See Fig.~\ref{fig:odd}. Therefore, this case can similarly be addressed by applying a Hadamard
test to estimate the inner product \(\bra{L}R\rangle\), analogously to
the estimation of the inner product
\(\bra{0}U_\psi^\dagger U_\phi\ket{0}\) in the even case. The states in Eq.~\eqref{LR} can be readily prepared given access to the
state-preparation input oracles for the matrices; indeed, their preparation
is even simpler than in the general even case, since no register
permutation is required.

For a multivariate trace of an odd number of general matrices,
\begin{equation}
    \operatorname{Tr}(A_1A_2\cdots A_{2k+1}),
\end{equation}
if one of the matrices can be decomposed as \(A_i=XY\), then the problem
reduces to the even case.
For example, if \(A_{2k+1}\) admits an efficient singular value
decomposition \(A_{2k+1}=U\Sigma V^\dagger\), then we can take
\(X=U\sqrt{\Sigma}\) and \(Y=\sqrt{\Sigma}V^\dagger\), so that
\[
    \|X\|_F\|Y\|_F=\|A_{2k+1}\|_1,
\]
where \(\|\cdot\|_1\) is the Schatten \(1\)-norm.
Let
\begin{equation}
    r_*
    =
    \frac{\|A_{2k+1}\|_1^2}{\|A_{2k+1}\|_F^2}
    \le
    \operatorname{rank}(A_{2k+1}).
\end{equation}
Then the multivariate trace estimation circuit needs to be run
\begin{equation}
    O\!\left(
    r_*F^2\epsilon^{-2}\log(1/\delta)
    \right)
\end{equation}
times, where \(F=\prod_{j=1}^{2k+1}\|A_j\|_F\).

\subsection{Treatment of observables}\label{appsec:O}
In the application to extract information about the eigenstate, the multivariate traces involve an observable $O$ given as a Pauli string. Since $O$ is unitary, we show that an input oracle for $O$ is unnecessary. Hence the algorithm does not introduce an additional norm factor in the
complexity.

If there is an odd number of matrices, the multivariate trace can always be
written in the form
\begin{equation}\label{tr_O}
    \operatorname{Tr}(A_1OA_2A_3\cdots A_{2k})
\end{equation}
by the cyclic property of the trace, with one of the \(A_i\) being the
density matrix \(\ket{\phi}\bra{\phi}\). Let \(u=\operatorname{vec}(A_1)\),
and let \(v\) and \(M\) be such that
\(Mv=\operatorname{vec}(A_{2k}^T A_{2k-1}^T\cdots A_2^T)\). Then we have
\begin{equation}
    \operatorname{Tr}(A_1OA_2A_3\cdots A_{2k})
    =
    u^T (O\otimes I) Mv.
\end{equation}
Construct $U_\phi$ and $U_\psi$ as in Lemma~\ref{mat_prod_vec} and define $U_\psi' = (I_{2^{2(k-1)q}} \otimes O \otimes I_{2^q})U_\psi$. Then 
\begin{equation}
    \frac{\operatorname{Tr}(A_1OA_2A_3\cdots A_{2k})}{\prod_{i=1}^{2k}\|A_i\|_F} = \bra{0^{\otimes 2kq}} U_\phi^\dagger U_\psi' \ket{0^{\otimes 2kq}}.
\end{equation}
Hence, the multivariate trace in Eq.~\eqref{tr_O} can be estimated by
applying a Hadamard test to \(U_\phi^\dagger U_\psi'\), as in
Theorem~\ref{trace_est}.

\begin{figure}
    \centering
    \includegraphics[width=0.25\linewidth]{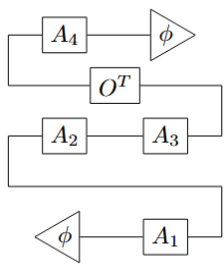}
    \caption{Tensor network representation of $\bra{\phi}A_1A_2A_3OA_4\ket{\phi}$.}
    \label{fig:odd_O}
\end{figure}

If there is an even number of matrices, e.g.,
\(\operatorname{Tr}(A_1A_2A_3OA_4A_5)\), with one of the \(A_i\), say
\(A_5\), being the density matrix \(\ket{\phi}\bra{\phi}\), we can define
states \(\ket{L}\) and \(\ket{R}\) as in Section~\ref{sec:odd} and insert
\(O\) in the appropriate position. Specifically, as illustrated in Fig.~\ref{fig:odd_O}, we define
\begin{equation}
    \begin{aligned}
        &\ket{R'} = \ket{\phi} \otimes ((O^T \otimes I)\vecket{A_3}) \otimes \vecket{A_1}, \\
        &\ket{L} = \vecket{\bar{A}_4} \otimes \vecket{\bar{A}_2} \otimes \ket{\phi}.
    \end{aligned}
\end{equation}
Then
\begin{equation}
    \langle L|R'\rangle = \frac{\operatorname{Tr}(A_1A_2A_3OA_4A_5)}{\prod_{i=1}^4 \|A_i\|_F} = \frac{\bra{\phi}A_1A_2A_3OA_4\ket{\phi}}{\prod_{i=1}^4 \|A_i\|_F}.
\end{equation}

\section{Proof of Theorem~\ref{parallel_eig}}\label{app:proof_eig}

\begin{proof}[Proof of Theorem~\ref{parallel_eig}]
To estimate an eigenvalue $\lambda$ of $H$ to the precision $\epsilon$, it suffices to estimate the eigenvalue $\lambda/h$ of $H/h$ to precision $\epsilon/h$. Let \(K = H/h,h\ge\|H\|_F\), and set the estimation precision to \(\hat{\epsilon} = \epsilon/h\). Let \(\lambda_1\) and \(\lambda_1^{\mathcal K}\) be the exact smallest
eigenvalues of \(K\) and \(K_D\), respectively.
Let \(\tilde{K}_D\) be the approximation of \(K_D\), where the entries of \(\tilde{K}_D\) are evaluated by the trace-estimation approach described above, and let \(\widetilde\lambda_1^{\mathcal K}\) be the smallest
eigenvalue of \(\tilde{K}_D\). 
By the triangle inequality,
\begin{equation}
    |\lambda_1-\widetilde\lambda_1^{\mathcal K}|
    \le
    |\lambda_1-\lambda_1^{\mathcal K}|
    +
    |\lambda_1^{\mathcal K}-\widetilde\lambda_1^{\mathcal K}|,
\end{equation}
so it suffices to ensure
\(|\lambda_1-\lambda_1^{\mathcal K}|\le \hat\epsilon/2\)
and
\(|\lambda_1^{\mathcal K}-\widetilde\lambda_1^{\mathcal K}|
\le \hat\epsilon/2\).

We invoke the error bound for the classical Lanczos method
\cite{doi:10.1137/0717059}:
\begin{equation}
    |\lambda_1-\lambda_1^{\mathcal K}|
    \le
    \frac{W\tan^2\theta_0}
    {\cosh^2((D-1)a)h},
\end{equation}
where $a = \mathrm{arcosh}\gamma$, $\gamma = 1+2\Delta/S_0$. Since
\(\cosh t=(e^t+e^{-t})/2\ge e^t/2\),
\begin{equation}
    |\lambda_1-\lambda_1^{\mathcal K}|
    \le
    4W\tan^2\theta_0\,e^{-2(D-1)a}/h.
\end{equation}
Thus it suffices to take
\begin{equation}
    D=1+
    \left\lceil
    \frac{1}{2a}
    \log\left(
    \frac{8W\tan^2\theta_0}{\epsilon}
    \right)
    \right\rceil.
\end{equation}

To guarantee
\(|\lambda_1^{\mathcal K}-\widetilde{\lambda}_1^{\mathcal K}|
\le \hat{\epsilon}/2\),
we require that $\delta = \frac{\hat{\epsilon}}{28\Gamma^2}$. Then, by Weyl's perturbation bound and by Lemma~\ref{lem:H_D},
\begin{equation}
    |\lambda_1^{\mathcal K} - \widetilde\lambda_1^{\mathcal K}|
    \le
    \|K_D-\tilde{K}_D\|_2
    \le
    14\delta\Gamma^2
    =
    \frac{\hat\epsilon}{2}.
\end{equation}
Since there are \(\Theta(D)\) multivariate traces to estimate (Eq.~\eqref{tr_momment}), the theorem then follows from the
complexity of multivariate trace estimation in Theorem \ref{trace_est} for 
\begin{equation}
    \delta=\min\left\{ \frac{\epsilon}{28h\Gamma^2}, \frac{1}{8D\Gamma^2}\right\}.
\end{equation}
The value of the failure-probability parameter is obtained from a
standard union bound argument.
\end{proof}

\begin{lemma}\label{lem:H_D}
    If 
    \begin{equation}
        \rho \coloneqq \delta D\Gamma^2 \le 1/8,
    \end{equation}
    then
    \begin{equation}
        \|\tilde{K}_D-K_D\|_2 \le 14\delta\Gamma^2.
    \end{equation}
\end{lemma}
\begin{proof}
    Let $\mu_l = \bra{\phi_0}K^l\ket{\phi_0},0\le l\le 2D-1$ and $\tilde{\mu}_l$ the estimations of $\mu_l$, and assume an uniform error bound 
    \begin{equation}
        \max\limits_{0\le l\le 2D-1} |\tilde{\mu}_l - \mu_l| \le \delta.
    \end{equation}
    Set \(e_l=\tilde{\mu}_l-\mu_l\) and interpolate the moments by \(\mu(t)=\mu_l+t e_l\) for \(0\le t\le 1\). Define 
    \begin{equation}
        G(t) = (\mu_{i+j}(t))_{i,j=0}^{D-1}, \qquad
        M(t) = (\mu_{i+j+1}(t))_{i,j=0}^{D-1}
    \end{equation}
    Note that $G(0) = X^\dagger X$ and $M(0) = X^\dagger KX$.
    We have 
    \begin{equation}
        \begin{aligned}
            & C^TG(0)C = I, \\
            & C^TG(t)C = I+tE, \qquad E = C^T(G(1)-G(0))C.
        \end{aligned}
    \end{equation}
    Since $\|E\|_2 \le \rho < 1/8$, $I+tE$ is positive definite for all $t \in [0,1]$. Let $S(t)$ be its upper-triangular Cholesky factor, with $S(t)^TS(t)=I+tE$. By the uniqueness of Cholesky factorization,
    \begin{equation}
        G(t) = C(t)^{-T}C(t)^{-1}, \qquad C(t) = CS(t)^{-1}.
    \end{equation}
    The corresponding perturbed Lanczos matrix is $K_D(t) = C(t)^TM(t)C(t)$.
    
    Let $\mathcal{L}_t$ be the polynomial functional with $\mathcal{L}_t(x^l)=\mu_l(t)$. We have the orthonormal conditions:
    \begin{equation}
        \mathcal{L}_t(p_i(t,\cdot)p_j(t,\cdot)) = \sum_{a,b} c_{i,a}(t)c_{j,b}(t)\mu_{a+b}(t) = \delta_{ij}.
    \end{equation}
    The matrix $K_D(t)$ is symmetric tridiagonal: If $j\ge i+2$, then $xp_i(t)$ has degree less than $j$, so $\mathcal{L}_t(xp_i(t)p_j(t))=0$. Its subdiagonal entries are positive because the leading coefficients of the polynomials are positive.
    Also, let $\mathcal{E}(x^l)=e_l$. For the Lanczos polynomials $p_j$,
    \begin{equation}\label{err_func}
        |\mathcal{E}(p_j)| \le \delta\|c_j\|_1 \le \delta \Gamma.
    \end{equation}
    Let $p_i(t,x)p_j(t,x) = \sum_a q_a(t)x^a$. Then $\mathcal{L}_t(p_ip_j) = \sum_a q_a(t)\mu_a(t)$. Note that $\dot{\mu}_l(t) = e_l$. By the chain rule,
    \begin{equation}
        \frac{\rm d}{\mathrm{d}t}\mathcal{L}_t(p_ip_j) = \mathcal{L}_t(\partial_t(p_ip_j))+\mathcal{E}(p_ip_j) = \mathcal{L}_t(\dot{p}_ip_j)+\mathcal{L}_t(p_i\dot{p}_j)+\mathcal{E}(p_ip_j).
    \end{equation}
    The derivative of $p_j$ admits the expansion $\dot{p}_j = \sum_{i \le j} \mathcal{L}_t(p_i\dot{p}_j)p_i$, i.e., $\mathcal{L}_t(p_i\dot{p}_j)$ is the coefficient of $\dot{p}_j$ in the basis vector of $p_i$. Differentiating the orthonormal conditions $\mathcal{L}_t(p_ip_j)=\delta_{ij}$ gives
    \begin{equation}\label{diff_p}
        \dot{p}_j = -\frac12\mathcal{E}(p_j^2)p_j - \sum_{i<j} \mathcal{E}(p_ip_j)p_i.
    \end{equation}
    Surpressing the argument $t$, we have
    \begin{equation}
        a_j = \mathcal{L}_t(xp_j^2), \qquad b_j=\mathcal{L}_t(xp_{j-1}p_j).
    \end{equation}
    Therefore,
    \begin{equation}
        \dot{a}_j = \mathcal{E}(xp_j^2)-a_j\mathcal{E}(p_j^2)-2b_j\mathcal{E}(p_{j-1}p_j).
    \end{equation}
    The final term is absent for $j=0$. Let $\kappa_j$ be the leading coeffiicient of $p_j$. $b_j = \kappa_{j-1}/\kappa_j$ by the recurrence relation Eq.~\eqref{poly_recur}. According to Eq.~\eqref{diff_p}, $\dot{\kappa}_j = -\frac12\mathcal{E}(p_j^2)\kappa_j$. We obtain
    \begin{equation}
        \frac{\dot{b}_j}{b_j} = \frac{\dot{\kappa}_{j-1}}{\kappa_{j-1}}-\frac{\dot{\kappa}_j}{\kappa_j} = -\frac12\mathcal{E}(p_{j-1}^2)+\frac12\mathcal{E}(p_j)^2.
    \end{equation}
    It follows that
    \begin{equation}
        \dot{b}_j = \frac{b_j}{2}[\mathcal{E}(p_j^2)-\mathcal{E}(p_{j-1}^2)].
    \end{equation}
    Note that $|a_j|,|b_j| \le 9/7$ by Lemma~\ref{lem:norm_bound}. Since $|\mathcal{E}(fg)| \le \delta\|\mathrm{coeff}(f)\|_1\|\mathrm{coeff}(g)\|_1$ and by Lemma~\ref{lem:1_norm}, we have
    \begin{equation}
        |\dot{a}_j| \le \frac{34}{7}\delta\Gamma_*^2, \qquad |\dot{b}_j|\le \frac{9}{7}\delta\Gamma_*^2
    \end{equation}
    Consequently, 
    \begin{equation}
        \|\dot{K}_D(t)\|_2 \le \|\dot{K}_D(t)\|_\infty \le |\dot{a}_j|+2|\dot{b}_j| \le \frac{52}{7}\delta\Gamma_*^2 \le 14\delta\Gamma^2.
    \end{equation}
    Integrating from $t=0$ to $t=1$ gives
    \begin{equation}
        \| \tilde{K}_D-K_D\|_2 = \|K_D(1)-K_D(0)\|_2 \le \int_0^1 \|\dot{K}_D(t)\|_2 \,\mathrm{d}t < 14\delta\Gamma^2.
    \end{equation}
\end{proof}

\begin{lemma}\label{lem:norm_bound}
    $\|K_D(t)\|_2 \le 9/7$.
\end{lemma}
\begin{proof}
    
    Set $F = C^T(M(1)-M(0))C$, then $K_D(t) = S(t)^{-T}(K_D+tF)S(t)^{-1}$.
    Since
    \begin{equation}
        \begin{aligned}
            & \|F\|_2 \le \delta\Gamma^2D = \rho, \qquad \|K_D\|_2 \le \|K\|_2 \le 1, \\
            & \|S(t)^{-1}\|_2^2 = \sigma_{\max}(S(t)^{-1})^2 = \frac{1}{\sigma_{\min}(S(t))^2} \le \frac{1}{\lambda_{\min}(S(t)^TS(t))} \le \frac{1}{1-t\rho},
        \end{aligned}
    \end{equation}
    we have
    \begin{equation}
        \|K_D(t)\|_2 \le \|S(t)^{-1}\|_2^2\|K_D+tF\|_2 \le \frac{1+t\rho}{1-t\rho} \le \frac{9}{7}.
    \end{equation}
\end{proof}

\begin{lemma}\label{lem:1_norm}
    $\|C(t)\|_1 \le \frac{4}{3}\Gamma \eqqcolon \Gamma_*$.
\end{lemma}
\begin{proof}
    For a matrix $A$, set $\|A\|_* = \max\{ \|A\|_1,\|A\|_\infty \}$. This norm is submultiplicative and invariant under tansposition. Set $Y(t)=S(t)-I$. Let $\Phi$ be a matrix map such that
    \begin{equation}
        (\Phi(A))_{ij} = \left\{
        \begin{aligned}
            &A_{ij}, i<j, \\ 
            &\frac12A_{ii}, i=j, \\
            &0,i>j.
        \end{aligned}
        \right.
    \end{equation}
    Then 
    \begin{equation}\label{eq:Y}
    Y(t) = \Phi(tE-Y(t)^TY(t)). 
    \end{equation}
    The map $\Phi$ does not increase $\|\cdot\|_*$. Hence $y(t)=\|Y(t)\|_*$ satisfies $y(t)\le t\rho+y(t)^2$. Since $y(0)=0$ and the Cholesky factor is continuous, $y(t)$ cannot reach; otherwise it would imply $1/2 \le 1/8+1/4$. Therefore,
    \begin{equation}
    \begin{aligned}
        &\|Y(t)\|_* = \|S(t)-I\|_* \le 2t\rho \le \frac{1}{4}, \\
        &\|S(t)^{-1}\|_1 \le \sum_{k=0}^\infty \|Y(t)\|_1^k \le \frac{4}{3}.
    \end{aligned}
    \end{equation}
    It follows that $\|C(t)\|_1 \le \frac{4}{3}\Gamma$.
\end{proof}

\section{Proof of Theorem~\ref{parallel_eigstate}}\label{app:eigstate}
We adopt the notation introduced in the proof of Theorem~\ref{parallel_eig}.

\begin{lemma}\label{lem:op_norm}
    For any normalized vector $v$, $\|(S^{-1}-I)v\|_2 \le 3\delta\Gamma^2\sqrt{D}\|v\|_1$.
\end{lemma}
\begin{proof}
    Set $m(t) = \max_{i,j}|(S(t)-I)_{ij}|$. Each entry of $E$ is at most $\delta\Gamma^2$ in absolute value. By Eq.~\eqref{eq:Y},
    \begin{equation}
        m(t) \le t\delta\Gamma^2+Dm(t)^2.
    \end{equation}
    Applying the same continuity argument to $Dm(t)$ as in Lemma~\ref{lem:1_norm} yields $m(t) \le 2t\delta\Gamma^2$. Since 
    \begin{equation}
        S(t)^{-1}-I = \sum_{n=1}^\infty (-1)^nY(t)^n
    \end{equation}
    and $\|S(t)-I\|_\infty \le 2t\rho$,
    \begin{equation}
        |(S(t)^{-1}-I)_{ij}| \le \sum_{n=1}^\infty |(Y^n)_{ij}| \le \sum_{n=1}^\infty\sum_k |Y_{ik}||(Y^{n-1})_{kj}| \le \sum_{n=1}^\infty 2t\delta\Gamma^2(2t\rho)^{n-1}.
    \end{equation}
    Hence, $m(t) \le \frac{2t\delta\Gamma^2}{1-2t\rho} \le \frac{8}{3}t\delta\Gamma^2$ and
    \begin{equation}
        \|(S^{-1}-I)v\|_2 \le m(1)\sqrt{D}\|v\|_1 = 3\delta\Gamma^2\sqrt{D}\|v\|_1.
    \end{equation}
\end{proof}

\begin{proof}[Proof of Theorem~\ref{parallel_eigstate}]
    Let $g \coloneqq \Delta/h$. Assume that the matrix moments $\mu_l$ and the observable moments $\Omega_{ij}$ are estimated with respective uniform additive errors
    \begin{equation}
        \delta_\mu = \min\left\{ \frac{\epsilon}{256\sqrt{D}\nu},\frac{1}{8\Gamma^2 D} \right\}, \qquad \delta_O = \frac{\epsilon}{8\nu}.
    \end{equation}
    Here $\nu = \Gamma^2\Lambda$ with $\Lambda = \max\{ V^2,g^{-1} \}$.
    Then 
    \begin{equation}
        \zeta \coloneqq \|\tilde{K}_D-K_D\|_2 \le 14\delta_\mu\Gamma^2 \le \frac{14\epsilon}{256\sqrt{D}\Lambda} \le \frac{\epsilon g}{16}.
    \end{equation}
    
    The proof analyzes the Ritz-vector error from $\tilde{v}_1$, the error from the noisy polynomial coefficients $\tilde{C}$, and the estimation error of observable moments $\tilde{\Omega}_{ij}$.
    \paragraph{Ritz-vector error.}
    Let $\Pi = PP^\dagger,w=(I-\Pi)\ket{\lambda_1},u=P^\dagger\ket{\lambda_1}$. Then $\ket{\lambda_1}=PP^\dagger\ket{\lambda_1}+w=Pu+w$. We have the identity
    \begin{equation}\label{ide}
        (\tilde{K}_D-\lambda_1I)u = (\tilde{K}_D-K_D)u-P^\dagger Kw.
    \end{equation}
    Then, by Weyl's perturbation bound, for $j>1$,
    \begin{equation}
        \tilde{\lambda}_j \ge \lambda_j-\zeta \ge \lambda_2-\zeta.
    \end{equation}
    Thus $\tilde{\lambda}_j-\lambda_1 \ge g-\zeta >0$ for $j>1$.
    Expand $u=\sum_{j=1}^D \nu_j \tilde{v}_j$ in terms of the eigenvectors of $\tilde{K}_D$ and let $Q=I-\tilde{v}_1\tilde{v}_1^\dagger$. Then
    \begin{equation}
        \begin{aligned}
        & Q(\tilde{K}_D-\lambda_1I)u = \sum_{j=2}^D (\tilde{\lambda}_j-\lambda_1)\nu_j\tilde{v}_j, \\
        & \|Q(\tilde{K}_D-\lambda_1I)u\|_2^2 = \sum_{j=2}^D |\tilde{\lambda}_j-\lambda_1|^2|\nu_j|^2 \ge (g-\zeta)^2 \sum_{j=2}^D |\nu_j|^2 = (g-\zeta)^2\|Qu\|^2.
        \end{aligned}
    \end{equation}
    By Eq.~\eqref{ide},
    \begin{equation}
        \|Q(\tilde{K}_D-\lambda_1I)u\|_2 = \|Q[(\tilde{K}_D-K_D)u-P^\dagger Kw]\| \le \|(\tilde{K}_D-K_D)u\|+\|P^\dagger Kw\|.
    \end{equation}
    It follows that
    \begin{equation}
        \|Qu\|_2\ \le \frac{\|(\tilde{K}_D-K_D)u\|+\|P^\dagger Kw\|}{g-\zeta}.
    \end{equation}
    Since $\|P^\dagger Kw\| = \|P^\dagger K(I-\Pi)w\| \le \|P^\dagger K(I-\Pi)\|\|w\|$, and $r \coloneqq \|(I-\Pi)K\Pi|| = \|\Pi K(I-\Pi)\| = \|P^\dagger K(I-\Pi)\| \le W/(2h)$, we have
    \begin{equation}
        \|Qu\|_2 \le \frac{r\|w\|_2+\zeta\|u\|}{g-\zeta}.
    \end{equation}
    Denote the Ritz-vector by $z = P\tilde{v}_1$ and write $u = \tilde{v}_1(\tilde{v}_1^\dagger u)+Qu$. Then
    \begin{equation}
        \ket{\lambda_1} = Pu+w = P\tilde{v}_1(\tilde{v}_1^\dagger u)+PQu+w = (\tilde{v}_1^\dagger u)z+PQu+w.
    \end{equation}
    Orthogonality of $PQu$ and $w$ implies that
    \begin{equation}
    \begin{aligned}
        \sin\angle(\ket{\lambda_1},z) 
        & = \|PQu+w\|_2 \\
        & = \sqrt{\|Qu\|^2+\|w\|_2} \\
        & \le \sqrt{\left( \frac{r\|w\|}{g-\zeta}+\frac{\zeta}{g-\zeta} \right)^2+\|w\|^2} \\
        & = \left\| \begin{pmatrix}
            \|w\| \\ \frac{r\|w\|}{g-\zeta}
        \end{pmatrix} + \begin{pmatrix}
            0 \\ \frac{\zeta}{g-\zeta}
        \end{pmatrix} \right\|_2 \\
        & \le \sqrt{1+\frac{r^2}{(g-\zeta)^2}}\|w\|_2+\frac{\zeta}{g-\zeta}.
    \end{aligned}
    \end{equation}
    By the convergence result of the classical Lanczos method \cite{doi:10.1137/0717059},
    \begin{equation}
        \|w\|_2 = \sin\angle(\ket{\lambda_1},\mathcal{K}_D) \le \frac{\tan\theta_0}{\cosh((D-1)a)}.
    \end{equation}
    Set $B = \sqrt{1+\frac{W^2}{\Delta^2}}$ and take
    \begin{equation}
        D = 1+\left\lceil \frac{1}{a}\log\left( \frac{16B\tan\theta_0}{\epsilon} \right) \right\rceil.
    \end{equation}
    Then $\|w\|_2 \le \frac{\epsilon}{8B}$. Since $\zeta \le \frac{\epsilon g}{16} < g/2$, the trace-distance bound for pure states yields
    \begin{equation}
        |\bra{\lambda_1}O\ket{\lambda_1}-z^\dagger Oz| \le 2\sin\angle(\ket{\lambda_1},z) \le 2B\|w\|_2+\frac{4\zeta}{g} \le \frac{\epsilon}{2}.
    \end{equation}

    \paragraph{Error from the noisy polynomial coefficients.}
    The practical Ritz-vector is
    \begin{equation}
        y = X\tilde{d} = X\tilde{C}\tilde{v} = XCS^{-1}\tilde{v} = PS^{-1}\tilde{v}.
    \end{equation}
    By Lemma~\ref{lem:norm_bound},
    \begin{equation}
    \begin{aligned}
        & \|y-z\|_2 = \|PS^{-1}\tilde{v}-P\tilde{v}\|_2 = \|(S^{-1}-I)\tilde{v}\|_2 \le 3\delta_{\mu}\Gamma^2\sqrt{D}V, \\
        & \|y\|_2 \le \|S^{-1}\|_2\le \sqrt{\frac{8}{7}}.
    \end{aligned}
    \end{equation}
    Consequently,
    \begin{equation}
        |y^\dagger Oy-Z^\dagger Oz| \le \|y-z\|_2(\|y\|_2+\|z\|_2)\le 7\delta_\mu\Gamma^2\sqrt{D}V \le \frac{7\epsilon V}{256\Lambda} \le \frac{\epsilon}{4}.
    \end{equation}
    The last inequality used $V \ge 1$ and $\Lambda \ge V^2$.

    \paragraph{Multivariate-trace estimation error.}
    Since $\|C\|_1 \le \Gamma$ and $\|S^{-1}\|_1 \le 4/3$,
    \begin{equation}
        \|\tilde{d}\|_1 \le \frac{4}{3}\Gamma V.
    \end{equation}
    Also $\tilde{d}^\dagger\Omega\tilde{d} = y^\dagger Oy$. We obtain
    \begin{equation}
        |\tilde{d}^\dagger(\tilde{\Omega}-\Omega)\tilde{d}| \le \delta_O\|\tilde{d}\|_1^2 \le \frac{16}{9}\delta_O\Gamma^2V^2 \le \frac{\epsilon}{4}.
    \end{equation}
    Taking the real part cannot increase this error. Adding the three contributions gives
    \begin{equation}
        \|\tilde{o}-o\| \le \epsilon.
    \end{equation}.

    There are $O(D)$ matrix moments and $O(D^2)$ observable moments. Their respective execution counts are
    \begin{equation}
        \begin{aligned}
            O(D\delta_\mu^{-2}\log(D/\eta)) &= O(D^2\max\{\nu^2\epsilon^{-2},D\Gamma^4\}\log(D/\eta)), \\
            O(D^2\delta_O^{-2}\log(D^2/\eta)) &= O(D^2\nu^2\epsilon^{-2}\log(D/\eta)).
        \end{aligned}
    \end{equation}
    A union bound gives total failure probability at most $\eta$. This proves the stated total complexity.
\end{proof}

\section{On the Errors in \cite{mor2025multivariate}\label{app_err}}

Theorem 17 in \cite{mor2025multivariate} is proved by mathematical induction. They assume that 
\begin{equation}
  \operatorname{Tr}(A_1\cdots A_{2k}) = \operatorname{vec}(A_1)^T F^{(\frac{k}{2})}(E_{1,k},\cdots,E_{k/2,k}) \operatorname{vec} \left( F^{(\frac{k}{2})}(O_{1,k},\dots,O_{k/2,k}) \right)
\end{equation}
for even $k$. In the inductive step, the authors show that
\begin{equation}\label{induction}
    \begin{aligned}
        \operatorname{Tr}(A_1\cdots A_{2k+2}) = 
        & \operatorname{vec}(A_1)^T (A_2 \otimes A_{2k+2}^T) \left( A_3 \otimes A_{2k+1}^T \otimes \operatorname{vec}\left(F^{(\frac{k-2}{2})}(O_{2,k+1},\dots,O_{\frac{k}{2},k+1})\right)^T \right) \\ 
        & \times \operatorname{vec}\left(F^{(\frac{k}{2})}(E_{2,k+1},\dots,E_{\frac{k+1}{2},k+1})\right).
    \end{aligned}
\end{equation}
They then place a trace around the scalar in Eq.~\eqref{induction} and invoke the identity $\operatorname{Tr}(ABCD) = \operatorname{vec(A)^T}(B\otimes D^T)\operatorname{vec}(C^T)$ with
$A=\operatorname{vec}(A_1)^T$, $B=A_2 \otimes A_{2k+2}^T$, $C=A_3 \otimes A_{2k+1}^T \otimes \operatorname{vec}\left(F^{(\frac{k-2}{2})}(O_{2,k+1},\dots,O_{\frac{k}{2},k+1})\right)^T$, and $D=\operatorname{vec}\left(F^{(\frac{k}{2})}(E_{2,k+1},\dots,E_{\frac{k+1}{2},k+1})\right)$. Applying this identity correctly yields
\begin{equation}
    \begin{aligned}
        \operatorname{Tr}(A_1\cdots A_{2k+2}) = 
        & \operatorname{vec}(A_1)^T \left( A_2 \otimes A_{2k+2}^T \otimes \operatorname{vec}\left( F^{(\frac{k}{2})}(E_{2,k+1},\dots,E_{\frac{k+1}{2},k+1}) \right)^T \right) \\
        & \times \operatorname{vec}\left( A_3^T \otimes A_{2k+1} \otimes \operatorname{vec}\left( F^{(\frac{k-2}{2})}(O_{2,k+1},\dots,O_{\frac{k}{2},k+1}) \right) \right),
    \end{aligned}
\end{equation}
which contradicts the induction hypothesis. However, the authors mistakenly obtain
\begin{equation}
    \begin{aligned}
        \operatorname{Tr}(A_1\cdots A_{2k+2}) = 
        & \operatorname{vec}(A_1)^T \left( A_2 \otimes A_{2k+2}^T \otimes \operatorname{vec}\left( F^{(\frac{k}{2})}(E_{2,k+1},\dots,E_{\frac{k+1}{2},k+1}) \right)^T \right) \\
        & \times \operatorname{vec}\left( A_3 \otimes A_{2k+1}^T \otimes \operatorname{vec}\left( F^{(\frac{k-2}{2})}(O_{2,k+1},\dots,O_{\frac{k}{2},k+1}) \right)^T \right),
    \end{aligned}
\end{equation}
rendering the induction step invalid.

In fact, Eqs.~\eqref{M} and~\eqref{v} indicate that $M$ and $v$ do not admit the same recursive construction, since $M$ involves transposes of vectorized matrices while $v$ does not. Therefore, the recurrence $F^{(p)}(X_1,\dots,X_p)=X_1 \otimes \operatorname{vec}(F^{(p-1)}(X_2,\dots,X_p))^T$ proposed in \cite{mor2025multivariate} for the construction of both $M$ and $v$ appears problematic.



\section{Distributing quantum fan-out operations}\label{app:fanout}

\begin{figure}
    \centering
    \includegraphics[width=0.45\linewidth]{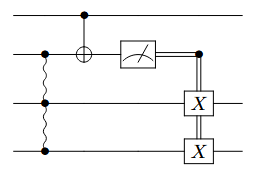}
    \caption{A cat-entangler. The squiggly line indicates that a generalized GHZ state is shared among the qubits marked by solid dots. The cat-entangler maps a single-qubit state on the first line, \(\alpha\ket{0}+\beta\ket{1}\), together with the generalized GHZ state
\(\frac{1}{\sqrt{2}}(\ket{000}+\ket{111})\), to the cat-like state
\(\alpha\ket{000}+\beta\ket{111}\).}
    \label{cat_entangler}
\end{figure}

The idea of Ref.~\cite{yimsiriwattana2004generalizedghzstatesdistributed} to implement non-local controlled gates is to employ a technique called cat-entangler to distribute control over multiple computers. A cat-entangler is shown in Fig.~\ref{cat_entangler}. It transforms an
arbitrary control-qubit state \(\alpha\ket{0}+\beta\ket{1}\), together with
a generalized GHZ state (also called a cat state, e.g.,
\(\frac{1}{\sqrt{2}}(\ket{000}+\ket{111})\)), into a cat-like state
(correspondingly, \(\alpha\ket{000}+\beta\ket{111}\)). Hence, we can use any qubit shared in the cat-like state as a control line.

\begin{figure}
    \centering
    \includegraphics[width=0.45\linewidth]{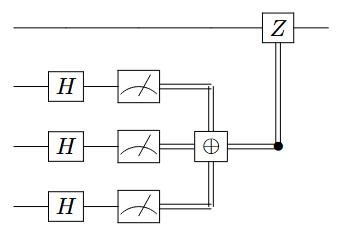}
    \caption{A cat-disentangler. Given the input state
\(\alpha\ket{0000}+\beta\ket{1111}\), the cat-disentangler outputs
\(\alpha\ket{0}+\beta\ket{1}\). The \(Z\) gate is controlled by the modulo-2 sum of the measurement outcomes \(r_k\), where \(r_k\) is the result of the measurement on line \(k\).}
    \label{cat_disentangler}
\end{figure}

\begin{figure}
    \centering
    \includegraphics[width=0.5\linewidth]{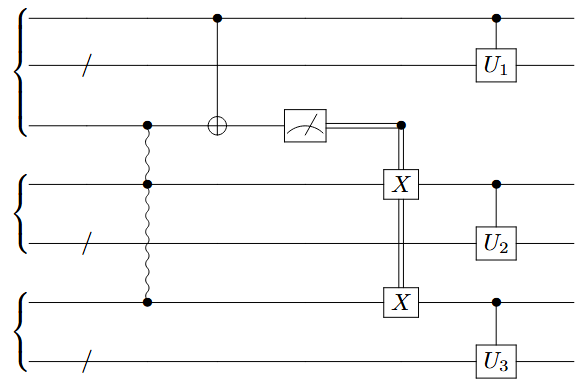}
    \caption{For \(U=U_1\otimes U_2\otimes U_3\), implementing the controlled-\(U\) operation with the cat-like state allows the control to be distributed among multiple computers and enables parallel execution on each computer. The qubits enclosed in curly braces are regarded as one sub-computer.}
    \label{distributed_cU}
\end{figure}

Assume that the cat state has been generated and shared among multiple
computers, and that \(U=U_1\otimes U_2\otimes U_3\) with each \(U_i\) acting
on a different computer. Figure~\ref{distributed_cU} then illustrates the
distributed implementation of the controlled-\(U\) operation.
Reference~\cite{yimsiriwattana2004generalizedghzstatesdistributed} points
out that the cat-like state can be used not only to distribute a control
line among multiple computers but also to execute controlled-\(U\) in
parallel. To complete the controlled-\(U\) operation, the control line must
be disentangled and restored, which can be accomplished by a
cat-disentangler, as shown in Fig.~\ref{cat_disentangler}. To distribute the
quantum fan-out, note that it can be viewed as a controlled-\(U\) operation
with \(U\) being a tensor product of gates.

\begin{figure}
    \centering
    \includegraphics[width=0.35\linewidth]{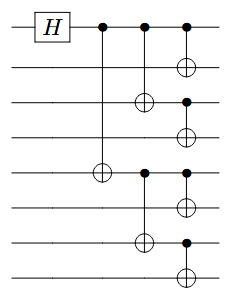}
    \caption{An efficient circuit for generating the cat state.}
    \label{cat_state}
\end{figure}

Generating and distributing an \(m\)-fold cat state consumes \(m-1\) entangled pairs (ebits) and \(2(m-1)\) classical bits (cbits) to implement \(m-1\) tele-CNOT gates, while requiring only \(O(\log m)\) steps as illustrated in Fig.~\ref{cat_state} \cite{yimsiriwattana2004generalizedghzstatesdistributed}. Both the cat-entangler and the cat-disentangler require \(m-1\) cbits. Consequently, a quantum fan-out across \(m\) modules can be distributed in $O(\log m)$ steps at a cost of \(m-1\) ebits and \(4(m-1)\) cbits.

\end{document}